\documentclass[sigconf]{acmart}
\newcommand{\tm}[3][]{{\color{black}\sout{#2}}\xspace{\color{blue}\uwave{#3}}\xspace{\ifx&#1&\else{\color{blue}[ts: #1]}\fi}}
\usepackage{comment}
\usepackage{graphicx}
\usepackage{adjustbox}
\usepackage{tikz}
\usepackage{pifont}
\usepackage{colortbl}
\usepackage{bm}
\usepackage{enumitem}
\usepackage{mdframed}
\usepackage{tcolorbox}
\usepackage{tabularx}
\usepackage{tabulary}
\usepackage{array}
\usepackage{setspace}
\allowdisplaybreaks
\usepackage{hyperref}
\usepackage{amsmath}

\usepackage{listings}

\usepackage{algorithm}
\usepackage{algorithmic}
\usepackage{amsmath}

\usepackage{tcolorbox}
\usepackage{refcount}

\newcounter{finding}

\usepackage{pifont}
\usepackage{subcaption}

\usepackage{listings}
\usepackage{xcolor}
\usepackage{multirow}
\usepackage{graphicx}
\usepackage[table]{xcolor} % これなんか悪さしないかな？

\usepackage{xspace}
\usepackage{pgfmath}
\usepackage{enumitem}
\newcommand{\sys}{SmartC2Rust\xspace}

\newenvironment{icompact}{
  \begin{list}{$\bullet$}{
    \itemindent -.05em
    \parsep 0pt plus 1pt
    \partopsep 0pt plus 1pt
    \topsep 2pt plus 2pt minus 2pt
    \itemsep 0pt plus 1.3pt
    \parskip 0pt plus 2pt
    \leftmargin 0.13in}
      }
{\normalsize\end{list}}

\copyrightyear{2026}
\acmYear{2026}
\setcopyright{cc}
\setcctype{by}
\acmConference[ICSE '26]{2026 IEEE/ACM 48th International Conference on Software Engineering}{April 12--18, 2026}{Rio de Janeiro, Brazil}
\acmBooktitle{2026 IEEE/ACM 48th International Conference on Software Engineering (ICSE '26), April 12--18, 2026, Rio de Janeiro, Brazil}
\acmDOI{10.1145/3744916.3773259}
\acmISBN{979-8-4007-2025-3/26/04}



\author{Momoko Shiraishi}
\affiliation{%
  \institution{The University of Tokyo}
  \city{Tokyo}
  \country{Japan}}
\email{shiraishi@os.is.s.u-tokyo.ac.jp} 

\author{Yinzhi Cao}
\affiliation{%
  \institution{Johns Hopkins University}
  \city{Baltimore}
  \country{USA}}
\email{yinzhi.cao@jhu.edu}

\author{Takahiro Shinagawa}
\affiliation{%
  \institution{The University of Tokyo}
  \city{Tokyo}
  \country{Japan}}
\email{shina@is.s.u-tokyo.ac.jp}

\begin{document}

\title[\sys]{
%LLM-based Translation from C to Compilable, Functional Equivalent, and Safe Rust Code
\sys: Feedback-Driven, Iteratively-Refined C-to-Rust Translation via Large Language Models for Safety and Semantic Equivalence}
\title[\sys]{\sys: Iterative, Feedback-Driven C-to-Rust Translation via Large Language Models for Safety and Equivalence}

%\thispagestyle{plain}

%\pagestyle{plain}

% \sys: Semantically-Equivalent, Safe C-to-Rust Translation via Feedback-Driven, Iterative Refinement of Outputs from Large Language Models

%via an Iterative,  


% \affiliation{%

% \affiliation{%

% \affiliation{%


\begin{abstract}

Memory safety vulnerabilities remain prevalent in today's software systems and one promising solution to mitigate them is to adopt memory-safe languages such as Rust.  Due to legacy code written in memory unsafe C, there is strong motivation to translate legacy C code into Rust.  Prior works have already shown promise in using Large Language Models (LLMs) for such translations.  However, significant challenges persist for LLM-based translation: the translated code often fails to compile, let alone reduce unsafe statements and maintain the semantic functionalities due to inherent limitations of LLMs such as limited token size and inconsistent outputs.

In this paper, we design an automated C-to-Rust translation system, called \sys, to segment and convert the C code to Rust with memory safety and semantic equivalence. The key insight is to iteratively refine the output Rust code with additional feedback, e.g., compilation errors, segmentation contexts, semantic discrepancies, and memory unsafe statements. Such feedback will gradually improve the quality of generated Rust code, thus mitigating unsafety, inconsistency, and semantic issues.  Our evaluation shows that \sys significantly decreases the unsafe statements and outperforms prior works in security and semantic equivalence.  %We also show that \sys is scalable to large-scale C codebase. 

\end{abstract}

\maketitle

% \definecolor{blue}{RGB}{0,0,0}

% Sections
\section{Introduction} \label{sec:intro}

Memory safety vulnerabilities, despite their long history, continue to pose significant threats to software systems written in unsafe languages such as C, leading to many critical security flaws~\cite{horizon3.ai}.  A promising solution for mitigating such vulnerabilities is to build software systems using memory-safe languages such as Rust.  Since many legacy software systems are still written in the C language, a natural research problem is translating legacy C code into Rust to enhance security~\cite{cisa2023}.

Existing translation approaches can be broadly categorized into two types: rule-based and large language model (LLM)-based. Rule-based approaches~\cite{c2rust,emre2021translating, hong2023concrat, emre2023aliasing, hong2024don}, e.g., C2Rust~\cite{c2rust}, rely on predefined translation mapping rules, such as rewriting raw pointers and unsafe blocks into safer Rust alternatives.
However, one-to-one conversion has limitations in accurately describing countless mappings and selecting the appropriate translations based on context. As a result, these approaches often still produce non-idiomatic and unsafe Rust code~\cite{pan2024lost}.
In contrast, LLM-based translations~\cite{eniser2024towards, yang2024vert, c2saferrust}, show promise in generating more idiomatic and safer Rust code with minimal reliance on unsafe constructs, potentially producing better, more secure code efficiently. However, LLM-based translation is still in its early stages and faces several challenges:

%~\cite{eniser2024towards, yang2024vert,c2saferrust}, e.g., C2SaferRust~\cite{c2saferrust}

% Several works on LLM-based C-to-Rust translation~\cite{eniser2024towards, yang2024vert} reported that it generates more idiomatic and safer Rust code with minimal reliance on unsafe constructs, offering the potential to produce better, more secure code efficiently.
% However, LLM-based translation is still in its early stages of development and faces several challenges.

%Challenges:

\begin{icompact}
\item Memory Safety. \hspace{0.05in} The translated Rust code needs to eliminate unsafe C features, instead of relying on unsafe blocks or raw pointers. However, prior works~\cite{c2saferrust, eniser2024towards} often fail to remove such features, undermining Rust’s memory safety improvements. 

\item Semantic Equivalence. \hspace{0.05in} The translated Rust code is expected to preserve the functionality of the original C code. However, existing approaches often fail to even compile or to pass the test cases accompanying the original C program, resulting in semantic inequivalence. 
Several prior works ~\cite{eniser2024towards, yang2024vert} have shown that LLMs struggle to generate compilable Rust code even small-scale C programs with fewer than 100 lines of code.
%This constraint also limits existing approaches, making it difficult to correctly translate even small-scale C programs with fewer than 100 lines of code into Rust~\cite{eniser2024towards, yang2024vert}.

\item Macro Translation. \hspace{0.05in} Translating C macros to Rust is challenging: Existing approaches often alter semantics by expanding macros in the C code for translation~\cite{pappas2024semantic}, which disables conditional compilation and reduces code portability across different systems or environments.

\item Limited LLM Context Windows~\cite{window}. \hspace{0.05in} Modern LLMs have a limited context window, restricting the number of tokens they can process at once. Previous studies~\cite{levy2024same, chen2023frugalgpt} have also shown that performance accuracy decreases as token count increases, even with 2,000-3,000 input tokens. 
%This constraint also limits existing approaches, making it difficult to correctly translate even small-scale C programs with fewer than 100 lines of code into Rust~\cite{eniser2024towards, yang2024vert}.
%LLMs can typically translate only small C programs, with fewer than 100 lines, into compilable Rust code~\cite{eniser2024towards, yang2024vert}.

\end{icompact}

In this paper, we design an automated C-to-Rust translation system, called \sys, to achieve memory safety and semantic equivalence under the constraint of limited LLM context windows.
The high-level idea is to iteratively refine the Rust code generated by LLMs with additional information related to memory safety or semantics until it passes the test.
Accordingly, \sys prompts the LLM to fix the generated Rust code based on feedback, such as compilation errors and differences in test case results between the C and Rust code.
To address the aforementioned challenges, \sys employs the following techniques: 

\begin{icompact}

\item Code Refinement for Improved Security. \hspace{0.05in} \sys detects unsafe translations by identifying insecure features such as raw pointers and unsafe blocks. It then prompts the LLM to refine the translation by eliminating these features.

\item Localization of Functional Error.  \hspace{0.05in} \sys runs both the original C code and the translated Rust code using test cases to validate semantic correctness. It compares execution traces, such as call stack traces and outputs like return values, to localize potential translation errors. \sys then prompts the LLM to fix these errors using the identified code snippets and observed output differences.  

%Additionally, 
\item Semantic Preservation of C Macros. \hspace{0.05in}
%Unlike rule-based approaches, 
 \sys preserves the semantics of C macros where possible by detecting and transforming them without expansion.
In addition to constants and functions defined with \texttt{\#define}, \sys transforms conditional compilation using \texttt{\#ifdef} blocks into Rust’s \texttt{cfg} attribute~\cite{cfg_attribute}, enabling the translated Rust code to support a single codebase without requiring developers to manage multiple separate versions for different configurations.

%\sys also detects and transforms C macros and enables transformation of C macros to Rust macros using LLMs, such as \texttt{\#ifdef} blocks with Rust’s \texttt{cfg} attribute. % for all the macros using LLMs. 

\item  Modular Segmentation with Translation Contexts. \sys segments C code into smaller fragments and supplements the translation of each fragment with relevant context, such as metadata from previous translations, to enhance consistency and accuracy. For example, \sys stores translated Rust function signatures along with their corresponding C function names and locations in a compressed database, ensuring consistent naming across translations of different fragments.

%refines the translation for 

%Even if the translated code compiles successfully, it must be functionally identical to the original C code.
%We define functional equivalence by whether test cases pass. If test cases fail, it is required to localize the code that need fixing. Unlike compile or runtime errors, functional errors may not output specific error messages and might simply result in different test case outcomes. For such case, we prepare the execution function flow of the original C test case as a \emph{golden flow}, and each time we run test cases with the translated Rust code, we output its execution flow as well. By comparing the current flow with the golden flow, we enable the LLM to localize the code that need correction, even if requiring repairs for relatively large programs.

\end{icompact}

% We plan to release an open-source version of \sys at this repository~\cite{repo} and an online demo at this URL~\cite{add}.
% We implemented a prototype of \sys at this repository~\cite{repo} and an online demo at this URL~\cite{add}.  
We provide artifacts of \sys at this repository~\cite{repo} and an online demonstration at this URL~\cite{add}.
 We compared \sys against state-of-the-art rule- and LLM-based translation tools, including C2Rust~\cite{c2rust}, Crown~\cite{zhang2023ownership}, Laertes~\cite{emre2023aliasing}, and C2SaferRust~\cite{c2saferrust}.  We evaluated \sys against \textcolor{black}{21} programs up to approximately 3,000 lines of code, and our evaluation shows that \sys outperforms prior works in reducing unsafe statements by \textcolor{black}{99.4}\%, improving test case pass rates by 11.25\%, and achieving \textcolor{black}{85.2} \% preservation of the semantic meaning of C macros. %We also evaluate the scalability of \sys and show that \sys can scale to a large-scale codebase with \todo{yy} Lines of code. 

\section{Overview} \label{sec:back}

In this section, we first describe a motivating example of C and translated Rust code with translation challenges, and then give an overview of our solution. 

\begin{figure}[t] 
\begin{lstlisting}[language=C, caption={An Example of Unsafe C Function}, label={lst:c_impl-1}]
   char *url_get_protocol(char *url) {
     char *protocol = malloc(URL_PROTOCOL_MAX_LENGTH * sizeof(char)); //allocate a memory
     if (!protocol) return NULL;
     sscanf(url, "%[^://]", protocol); //risk of overflow
     if (url_is_protocol(protocol)) return protocol;
     return NULL; 
   }
\end{lstlisting}
%\begin{figure}[t] 
\begin{lstlisting}[language=C, caption={Rust Code Translated with \sys.}, label={lst:rust_impl-1}, basicstyle=\ttfamily\footnotesize]
   use url::Url;
   pub fn rust_url_get_protocol(url: &str) -> Option<String> {
    let parsed = Url::parse(url).ok()?; // parsing
    let protocol = parsed.scheme().to_string();
    if rust_url_is_protocol(&protocol) {
        return Some(protocol);
    }
    None
   }
\end{lstlisting}
%\end{figure}
\end{figure}

\begin{figure}[t] 
\begin{lstlisting}[language=C, caption={Rust Code Translated with C2Rust.}, label={lst:rust_impl-2}, basicstyle=\ttfamily\footnotesize]
   #[no_mangle]
   pub unsafe extern "C" fn url_get_protocol(
       mut url: *mut libc::c_char,
   ) -> *mut libc::c_char {
    let mut protocol: *mut libc::c_char = malloc(
    (16 as libc::c_int as libc::c_ulong)
     .wrapping_mul(::core::mem::size_of::<libc::c_char>() as libc::c_ulong),
    ) as *mut libc::c_char;
    if protocol.is_null() {
        return 0 as *mut libc::c_char;
    }
    sscanf(url, b"%[^://]\0" as *const u8 as *const libc::c_char, protocol); //still uses sscanf()
    if url_is_protocol(protocol) {
        return protocol;
    }
    return 0 as *mut libc::c_char;
   }
\end{lstlisting}
\end{figure}

\subsection{A Motivating Example}

We present a motivating example in \autoref{lst:c_impl-1}, an unsafe C function, \texttt{url\_get\_protocol()}, from the real-world program \texttt{urlparser}~\cite{emre2023aliasing}.
The function extracts the protocol name, e.g., ``http,'' ``https,'' and ``ftp,'' from a URL string into \texttt{char *protocol}.
It then returns the protocol if the extracted name is valid or \texttt{NULL} otherwise.
The function has two memory safety issues.  
First, it allocates a fixed-size memory buffer at Line 2, but \texttt{sscanf()} at Line 4 can copy an arbitrarily long protocol name into the \texttt{protocol} buffer, exceeding the allocated size and causing a heap buffer overflow.
Second, the allocated memory buffer may not be freed, leading to a memory leak and potentially leaving a dangling pointer.

\autoref{lst:rust_impl-1} shows the correctly translated Rust code of this function using \sys.  
First, \texttt{sscanf()} is replaced with the \texttt{parse()} method from Rust's standard \texttt{Url} parsing library~\cite{std_url}, eliminating the heap buffer overflow while preserving the original URL parsing functionality.  
Second, unused data is automatically deallocated through Rust’s ownership and borrowing mechanism, preventing memory leaks and other safety issues.  
Thus, the translated Rust code successfully eliminates both memory safety issues in the original C implementation while maintaining semantic equivalence.  

%\texttt{parse()} method safely handles URL parsing and no manual memory allocation or buffer management is required. 

Such a C-to-Rust translation, albeit intuitively simple, is challenging for both rule-based and LLM-based approaches from the following three aspects.  
First, translating a C program into Rust code without using \texttt{unsafe} is not straightforward.  
For example, C2Rust~\cite{c2rust}, a state-of-the-art rule-based translator, converts the C code in \autoref{lst:c_impl-1} into Rust code in \autoref{lst:rust_impl-2}.  
This code retains \texttt{sscanf()} at Line 12 within an \texttt{unsafe} block and also uses \texttt{malloc()} along with a Rust raw pointer to manage the \texttt{protocol} buffer, failing to address any memory safety issues. Similarly, C2SaferRust~\cite{c2saferrust}, an existing LLM-based translator, generates code that calls C functions within an \texttt{unsafe} block.  
This occurs because C2SaferRust relies on C2Rust for initial translation and then applies an LLM for refinement, but the LLM fails to eliminate the \texttt{unsafe} code.  

Second, maintaining semantic equivalence before and after translation is challenging.  
For example, the URL parsing functionality is implemented using \texttt{sscanf()} and \texttt{url\_get\_protocol()} in the original C code, whereas the Rust code uses the corresponding \texttt{Url::parse()} method and \texttt{rust\_url\_is\_protocol()}.  
Correctly translating such functions requires a deep understanding of the semantics of both the original C code and the translated Rust code.
As with human developers, fully understanding the semantics solely from the textual representation without executing them is not easy.  

Third, maintaining translation consistency across the entire program is challenging.  
As previously mentioned, the LLM context window size limitation prevents the entire program from being translated at once.  
Thus, the translation process is divided into smaller units, such as the code shown in \autoref{lst:c_impl-1}.  
However, this often leads to inconsistencies between different translation units.  
For example, when generating code to call another function that has already been translated in a separate unit, LLMs may fail to retain information from previous translations, resulting in inconsistencies in names, functionality, or other aspects.

\subsection{Overall Solution}

We now provide an overview of \sys and describe how it addresses the three challenges using the same example.

First, \sys ensures memory safety during translation through iterative refinement.
LLMs often generate unsafe code even when explicitly instructed not to use it.
Thus, \sys first identifies unsafe patterns, such as \texttt{unsafe} blocks and raw pointers, as seen in \autoref{lst:rust_impl-2}, in each translation.  
It then prompts the LLM to refine the code until such patterns reduce. %no longer exist.  

Next, \sys ensures semantic equivalence through execution result comparison.
That is, \sys executes both the original and translated code using test cases accompanied with the original C code.  
Then, it compares the execution results, including the call stack leading to the function, the inputs (e.g., the \texttt{url} variables at Line 1 of \autoref{lst:c_impl-1} and Line 2 of \autoref{lst:rust_impl-1}), and the return values (e.g., Line 5 of \autoref{lst:c_impl-1} and Line 6 of \autoref{lst:rust_impl-1}).  
If the input and output of the entry point function are identical, \sys considers them semantically equivalent. 
%If all these match across test cases, \sys considers them semantically equivalent.  
Otherwise, \sys prompts the LLM to refine the translated code based on the identified differences and the related function.
%, i.e., \texttt{rust\_url\_get\_protocol()}.  
Note that, while \sys tracks the input and output of all functions as supplementary information, it judges semantic equivalence using only the entry point function rather than verifying unit tests for every function, recognizing that C and Rust functions may not have a one-to-one correspondence. For example, for standalone programs, we implement end-to-end testing to verify semantic equivalence.

Finally, \sys ensures consistency between translations by maintaining supplementary correspondence metadata.  
It first reorders code elements so that closely related code is placed together, making it easier to maintain context.  
Next, during translation, it generates summary information, such as function signatures and data types, alongside the target C code to improve consistency.  
For example, \sys provides the \texttt{url\_is\_protocol()}'s function signature %, i.e., \texttt{fn rust\_url\_is\_protocol(scheme: \&str) -> bool} 
 in translating the code in \autoref{lst:c_impl-1} so that the translation can correctly capture the dependency.

\begin{figure*}[t]
    \centering
    \includegraphics[width=0.75\textwidth]{ov.pdf}  % width=\linewidth
    \caption{Overview of \sys.} \label{fig:arch}
    \label{overview}
\end{figure*}
% \todo{Momoko: (1) Add bigger boxes for three stages, and (2) maybe merge pre-processing with segmentation; (3) do we still have a correspondence-mapping step? I do not see it in the description.  Maybe we can remove it.  Then,  we have five steps instead of seven?}

\section{Design}

In this section, we describe the design of \sys from its system architecture and three major stages.

\subsection{System Architecture}

Figure~\ref{fig:arch} shows the architecture of \sys, which has three major stages: (i) pre-processing, (ii) translation and quality check, and (iii) feedback-driven code repair.  Our high-level idea is to break down code into smaller parts with contexts from previous translations so that LLMs can provide more effective responses. 

%Our consistent policy across all steps is to break tasks into smaller parts so that the LLM can provide more effective responses.

In Stage (i), \sys merges C code from multiple files into larger modules, reorders code elements, and segments the merged code into smaller translation units.  
It also parses each unit to generate metadata, later used in prompts to supplement translation context across runs.  
In Stage (ii), \sys translates each C unit into a corresponding Rust unit using LLMs, supplementing prompts with metadata from both C and Rust as context.  
It then refines the translated Rust code to improve idiomaticity.  
In Stage (iii), \sys repairs the translated Rust code based on feedback from compilation or test cases.

\subsection{Stage (i): Pre-processing}

The goal of Stage (i) is to transform C code for better LLM comprehension.
It consists of three steps:  
(1) merging C code into modules for information aggregation,  
(2) code reordering and macro transformation to provide translation context, and  
(3) C module segmentation to fit LLM token limits.  

\subsubsection{Merging C code into modules}

\sys analyzes the C target program to detect the locations of function declarations and definitions, and then  % ctags~\cite{ctags}
%For macro variables and macro functions, we use a custom parser.
%We then 
merges them by copying the contents of the included files into those containing the \texttt{\#include} directives.
To avoid conflicts, \sys ensures that the namespaces of static variables and functions are unique.  
It also allows LLMs to reference both declarations and definitions within their context windows, facilitating the generation of accurate Rust code.
Since Rust does not separate declarations into different files like C, the translated Rust code can be used directly as a Rust module.

The rationale behind this merging step is to unify how functions and types are referenced across files in C and Rust.
In C, declarations are typically separated from definitions: header files contain function and type declarations, while source files hold their definitions and implementations.
In contrast, Rust uses the \emph{use} keyword to import modules, with functions and variables both declared and defined within a single module file.
The merging step consolidates declarations and definitions in the C source code, facilitating subsequent translation to Rust.

\subsubsection{Code Reordering and Macro Transformation}

This step consists of two sub-steps.  
First, \sys reorders code elements, such as functions and type definitions, in the merged module in topological order---that is, they appear before being referenced.
This improves name resolution and module importing, as all functions and types are already defined in C during translation and such definitions can be provided as contextual metadata.

Second, 
 \sys detects all macros for conditional compilation in C without expanding them, 
 %transforms all \texttt{\#ifdef} blocks with Rust’s \texttt{cfg} attribute,
  and then asks the LLM to determine whether the macro should be defined in \texttt{Cargo.toml}, \texttt{build.rs}, \texttt{lib.rs}, or each module file. %record and remove it from the source code, and later enable it in the build script (\texttt{build.rs}).
 This analysis enables the C macros to be semantically preserved in the translated Rust code. For example, \texttt{\#ifdef} blocks are detected and predetermined to be used with the \texttt{cfg} attribute, and this analysis result is used in translation.
 %Next, \sys enables and defines these macros in the corresponding files. 
 Note that macros are traditionally challenging in translation, because they are processed by a preprocessor rather than a compiler.  C2Rust, a prior rule-based translation, expands all macros into C, which may lead to semantic difference for different environments if \texttt{\#ifdef} is present.

\subsubsection{C Module Segmentation}

This step consists of (1) segmentation and (2) contextual metadata extraction.
First, \sys segments C code modules into smaller translation units.
It includes as much code as possible in each unit while preserving semantics to improve translation efficiency.
Specifically, \sys maintains a counter to track included code and divides modules based on \emph{logical blocks}.
A logical block refers to a semantically cohesive unit in C, such as function or type definitions, with examples listed in \autoref{tab:logicblock}.
\sys selects the closest logical block boundary to a given line of code and segments the C module accordingly.
Functions with circular dependencies are treated as a single logical block to ensure LLMs can resolve function references correctly.

Second, \sys analyzes the original C translation units to extract contextual metadata.  
Specifically, it records the definitions and call sites of code elements in the source code, storing this data in JSON format.  
This metadata supplements context between translation units without overloading the context window, thereby improving translation accuracy.

% When C code is segmented into translation units and each unit is translated independently by the LLM, the context between the units is lost.
% %To address this, we add information to supplement the missing context, allowing the LLM to translate from C to Rust effectively.
% %Given the size limitations of the context window, it is crucial that this context information be as small and concise as possible.
% To achieve this, we analyze the original C translation units in advance to extract contextual metadata and later add information to supplement the missing context.
% % Specifically, we classify C code into six elements:
% % \begin{enumerate}
% % \item functions
% % \item macro functions
% % \item type definitions
% % \item macro variables
% % \item global variables
% % \item others (the rest of the above categories such as header inclusion statements) % (see \autoref{code_element}).
% % \end{enumerate}
% For each element within each block of a unit, we record the definition and call sites of each code element in the source code, storing this data in JSON format as contextual metadata.

% When translating each segmented translation unit sequentially, we add this summarized contextual metadata to the prompt.
% This approach supplements the context between translation units without overloading the context window, thereby improving translation accuracy.

\subsection{Stage (ii): Translation and Quality Check}  

In Stage (ii), \sys translates each C unit and refines the translated Rust code.

\subsubsection{Translation} 
This step translates a given C unit into Rust using the translation prompt shown in \autoref{fig:example-prompt}.
The prompt consists of four parts: translation rules, contextual metadata, split responses, and C source code.
The contextual metadata is extracted during segmentation in Stage (i) and updated in Stage (ii).
Specifically, it is enriched using another LLM prompt that includes both the C and translated Rust code to incorporate Rust function signatures.
This enhancement of contextual metadata improves the quality and consistency of the Rust translation.

The response split handles the LLM output token size limit.  
\sys instructs the LLM to split the response into multiple parts if it exceeds the length limit.  
It preserves context from earlier parts and merges all responses into one.  

%For split responses, we design our system so that the context from earlier parts of the answer is preserved. We instruct all responses to be returned in JSON format.

\begin{table}[!t] \footnotesize
\caption{Logic Block Definitions and Examples} \label{tab:logicblock}
\centering
\begin{tabular}{lc}
\toprule
\textbf{Logic Blocks} & \textbf{Example} \\
\midrule
 Functions & int add(int a, int b) \{ return a + b; \} \\
 Global Variables & static char buffer[BUFFER\_SIZE]; \\
 Type Definitions & typedef struct \{ int x, y; \} Point; \\
 Macro Functions & \#define MAX(a, b) ((a) > (b) ? (a) : (b)) \\
 Macro Variables & \#define BUFFER\_SIZE 1024 \\
 Preprocessor Conditionals & \#ifdef DEBUG ... \#endif \\
 Header Inclusions & \#include <stdio.h> \\
\bottomrule
\end{tabular}
%\todo{Momoko: Please fill in the table.  Thanks!}
\end{table}

\subsubsection{Quality Check}  
\label{self_check}  

This step refines the translated Rust code to improve quality.  
The prompt in this step consists of three parts: (1) Rust code, (2) refinement guidelines, and (3) self-evaluation rules.  
The refinement guidelines define rules for enhancing Rust code quality, such as removing \texttt{unsafe} blocks and raw pointers and avoiding \texttt{static mut} for global variables.  
The self-evaluation flag ensures that LLMs adhere to the instructions.  
The translation process repeats until all flags are set to true.  
The flags include:
\begin{icompact}
\item \textbf{"refined\_completed"}: The Rust code has been successfully refined to proper Rust style, improving safety and adherence to Rust principles. 
\item \textbf{"current\_block\_complete"}: The Rust code fully implements the functionality of the original C segment without mocking, simplifications, or placeholders.
\item \textbf{"no\_omission"}: The Rust code contains no omissions and can be executed as-is.  
\end{icompact}

% LLMs often tend to omit or simplify content in their responses, so we need the latter two feedback to ensure no such code exists.
% We also require responses to include a flag "unsafe\_used", which indicates whether unsafe code was included, but by default we do not use this flag as a condition for repeating the process. The reason for this will be discussed in the results section.

\begin{figure}[!t]
\centering
\footnotesize
% Left column (top and bottom)
    % Top-left figure (new)
    \begin{mdframed}[innerleftmargin=5pt, innerrightmargin=5pt, innertopmargin=5pt, innerbottommargin=0pt]
    \footnotesize %\small 
%    \textsf{Please convert the following C language source code into Rust without using unsafe. During the conversion process, please strictly apply all the specified rules below.\\ 
    \textsf{Translate the following C code into Rust by strictly following the rules below.\\ 
    \textcolor{red}{\#\# Translation rules}\\
%    - To allow importing from other modules, declare all items (structures, enums, functions, constants, etc.) using pub (public).\\
    - Declare all items (structures, enums, functions, constants, etc.) using pub (public) to allow importing.\\
%    - When calling external functions, avoid using unsafe. Whenever possible, use Rust's standard library or crates to achieve the same functionality in a safe way.\\
    - Avoid unsafe in external function calls by using safe equivalents from Rust’s standard libraries or crates.\\
    \textcolor{red}{\#\# Contextual Metadata}\\
    - The definitions of the elements used in the C code have been provided in Rust in other modules as follows.\\
    \,\,\, $\cdot$ pub fn quadtree\_search(tree: \&Quadtree, x: f64, y: f64) -> Option<\&QuadtreePoint>  (defined in crate::quadtree\_0)\\
    \,\,\, $\cdot$ pub fn quadtree\_node\_isleaf(node: \&QuadtreeNode) -> bool (defined in crate::quadtree\_0)\\
    ...\\
    \textcolor{red}{\#\# Split Response}\\
    - If the answer exceeds the token size ...\\
    \textcolor{red}{\#\# C source code:}\\
    \texttt{[}Paste the C unit code\texttt{]}\\
    }
    \end{mdframed}
    \caption{An Example Translation Prompt} \label{fig:example-prompt} % \caption{An Exemplary Prompt for the C-to-Rust Translation}
    %\captionof{figure}{An example prompt in the translation.}
    \label{pro_conv}

\end{figure}

\subsection{Stage (iii): Feedback-driven Code Repair}
%\begin{figure}[t]
%\centering
\begin{algorithm}[!t]
%\scriptsize
\caption{Feedback-driven Code Repair} 
\label{alg:conversion} 
\begin{algorithmic}[1]
\scriptsize
\REQUIRE Input: $C\_UNIT_1, C\_UNIT_2, \ldots, C\_UNIT_n$, $Dep_C$: \text{C metadata} 
\ENSURE Output: $S = \{R\_UNIT_0, R\_UNIT_1, R\_UNIT_2, \ldots, R\_UNIT_{n}\}$%(a set of translated Rust code)
%\STATE $Dep_C \leftarrow \texttt{parse}(S)$
\STATE $S \leftarrow \emptyset$ \text{, where S: Translated Rust program}
\STATE $COMPILE\_MAX \leftarrow N$
\STATE $REPAIR\_MAX\leftarrow N'$
\FOR{$i \leftarrow 0$ \TO $n$}
    \IF{$i = 0$}
        \STATE $R\_UNIT_0 \leftarrow \texttt{translate\_state\_macros}(Dep_C)$  %\COMMENT{Convert dependencies to Rust file}
    \ELSE
        \STATE $R\_UNIT_i \leftarrow \texttt{translate}(C\_UNIT_i, Dep_C)$  %\COMMENT{Convert C file to Rust file with dependencies}
        \WHILE{\text{not all evaluation flags are true}}
            \STATE $R\_UNIT_i \leftarrow \texttt{refine\_quality}(R\_UNIT_i)$ 
        \ENDWHILE
    \ENDIF
    %\STATE $R\_UNIT_i \leftarrow \texttt{convert}(C\_UNIT_i, Dep_C)$  %i=の時 %\COMMENT{Convert C file to Rust file}
    \STATE $S \leftarrow S \cup \{R\_UNIT_i\}$  %\COMMENT{Add translated Rust file to the set $S$}
    \STATE $error \leftarrow \texttt{compile}(S)$  %\COMMENT{Compile the set $S$ and get errors}
%\STATE $error \leftarrow \texttt{compile}(S)$  %\COMMENT{Compile the set $S$ and get errors}
    \STATE $compile\_count \leftarrow 0$
    \WHILE{$compile\_error \neq \emptyset$ \AND $compile\_count < COMPILE\_MAX$}
        \STATE $Dep_R \leftarrow \texttt{parse}(S)$ \text{, where $Dep_R$: Rust metadata}
        \STATE $M \leftarrow \texttt{ask\_repair}(S, Dep_R)$ \text{, where $M$: Modification set} %\COMMENT{Get the modifications needed to fix the errors}
        %\STATE $S \leftarrow \texttt{reflect\_modifications}(S, M)$ 
        \FOR{$(R\_UNIT, revision\_content) \in M$}  % \COMMENT{For each modification in the set}
            \STATE $S \leftarrow \texttt{apply\_modification}(S, R\_UNIT, revision\_content)$  % \COMMENT{Apply the modification}
        \ENDFOR
        \STATE $compile\_error \leftarrow \texttt{compile}(S)$  %\COMMENT{Recompile $S$ to check for errors}
        \STATE $compile\_count \leftarrow compile\_count + 1$
    \ENDWHILE
    \FOR{$R\_UNIT \in M$} 
        \STATE $ Dep_C \leftarrow  \texttt{map\_correspondence}(R\_UNIT, C\_UNIT, Dep_C)$  % 変更のあったRファイルに対応するCファイルについて、corresponseを聞く
    \ENDFOR
\ENDFOR
\texttt{}
\WHILE{$compie\_error \neq \emptyset$ \AND $semantic\_error \neq \emptyset$ \textcolor{black}{\AND $runtime\_error \neq \emptyset$} \AND $repair\_count < REPAIR\_MAX$}
    \STATE $M \leftarrow \texttt{ask\_repair}(S)$  %\COMMENT{Get the modifications needed to fix the errors}
    %\STATE $S \leftarrow \texttt{reflect\_modifications}(S, M)$ 
    \FOR{$(R\_UNIT, revision\_content) \in M$}  % \COMMENT{For each modification in the set}
        \STATE $S \leftarrow \texttt{apply\_modification}(S, R\_UNIT, revision\_content)$  % \COMMENT{Apply the modification}
    \ENDFOR
    \STATE $compile\_error \leftarrow \texttt{compile}(S)$  %\COMMENT{Recompile $S$ to check for errors}
    \STATE $semantic\_error\textcolor{black}{, runtime\_error} \leftarrow \texttt{run\_test}(S)$  %\COMMENT{Recompile $S$ to check for errors}
    \STATE $repair\_count \leftarrow repair\_count + 1$
\ENDWHILE

\RETURN $S$
\end{algorithmic}
\end{algorithm}
% \caption{Translation Procedure}
% \label{alg:conversion}
% \end{figure}

Stage (iii) repairs the Rust code output from Stage (ii) by addressing syntactic and semantic errors.  
\autoref{alg:conversion} outlines this repair process, integrated with translation and quality refinement (Lines 5-–12).
In the algorithm, $C\_UNIT$ and $R\_UNIT$ denote a C or Rust translation unit, respectively, and $S$ represents the translated Rust program. 
\sys iterates through all errors and attempts repairs until none remain (Lines 13--28 for compilation repair and 29--37 for semantic repair).  
We now describe two key steps in the repair process: (1) error collection and (2) iterative repair.

\subsubsection{Error Collection}

\textcolor{black}{
\sys repairs four types of errors: 
\begin{icompact}  
\item Syntactic Errors during Compilation. \hspace{0.05in} \sys compiles the translated Rust code and records compilation errors (\emph{compile\_error} in \autoref{alg:conversion}). Raw compile error logs are directly included for later repairs. 
\item Syntactic Errors during Runtime. \hspace{0.05in} \sys runs test cases through the translated Rust code and records runtime errors (\emph{runtime\_error} in \autoref{alg:conversion}). Examples of such errors are like 
 array index out of bounds leading to a panick. 
\item Runtime Behavior Differences. \hspace{0.05in} \sys runs test cases on both the original C and translated Rust code to capture differences in runtime behavior. These discrepancies are represented as function execution flow information, as shown in \autoref{fig:semanticerror}.  
\item Semantic Errors during Runtime. \hspace{0.05in} \sys runs test cases through the translated Rust code and records semantic (assertion) errors (\emph{semantic\_error} in \autoref{alg:conversion}). Assertion error logs are directly included for later repairs.  
\end{icompact} 
% The last three error types can be illustrated as follows: Syntactic errors during runtime occur when the translated Rust code crashes during execution, such as array index out of bounds that produces thread `main' panicked at `index out of bounds: the len is 5 but the index is 10'. Runtime behavior differences capture cases where C and Rust behave differently, such as the intermediate state of execution functions is different with the input argument equals to 3 and the input argument equals to None. Semantic (assertion) errors during runtime happen when test cases fail due to incorrect test output results, such as a test returning 7 instead of the expected value 6.
}
The compilation repair (Lines 13--28) fixes compilation errors after each unit. The semantic repair (Lines 29--37) fixes all error types after all units are translated and tested.
\begin{comment}
In the comilation repair process that runs after each translation unit is translated and compiled (Lines 13--28), simply compilation errors are fixed. In the semantic repair process that runs after all units have been translated and test cases are executed (Lined 29--37), both types of errors are fixed. 
\end{comment}

% end-to-end tests, which also incorporate runtime error verification.
% %and functional equivalence errors requires more than just repeatedly running the program. Writing unit tests for every function, including setting up mocks and stubs, is time-consuming, so we do not verify these errors at each individual code segment. 
% However, since it is effective for LLMs to handle smaller tasks, for fixing runtime and functional errors, we attempt to break the task into more manageable parts by leveraging function execution flow information.

% As shown in the flow information snippet below, we maintain expected (golden) flow information from running the original C program, alongside function flow information from the Rust program. This Rust flow information is updated each time the LLM makes modifications and functions as one feedback. The LLM uses this 

\begin{figure}
\begin{lstlisting}[language=C, label={}, basicstyle=\ttfamily\footnotesize, numbers=none]
[Expected C Program Flow]
$\blacktriangleright$ main (which.c:50)
   IN $\rightarrow$ (argv:["./which", "--help"])
   $\blacktriangleright$ uidget (bash.c:30)
      OUT $\leftarrow$ 0
   $\blacktriangleright$ path_search (which.c:20)
   IN $\rightarrow$ (cmd: --help, path_list: ...) // ... continues

[Current Rust Program Flow]
$\blacktriangleright$ rust_main (module1/unit0.rs:40)
   IN $\rightarrow$ (args:["./which", "--help"])
   $\blacktriangleright$ rust_uidget (module4/unit0.rs:35)
      OUT $\leftarrow$ 0
   $\blacktriangleright$ rust_path_search (module4/unit0.rs:120)
   IN $\rightarrow$ (cmd: --help, path_list: ...) // ... continues
\end{lstlisting}
\caption{An example of function execution flow. } \label{fig:semanticerror}
\end{figure}

% $\vdots$ // execution continues

\subsubsection{Iterative Repair}

The repair prompt contains the following information: (1) response guidelines (i.e., rules for the repair), (2) Rust Code, (3) syntactic or semantic errors, (4) directory structure, and (5) dependencies (i.e., definitions of elements in other modules). 
Additionally, for semantic repair, the prompt includes (6) function execution flow information and (7) C code.

Each repair may span over multiple Rust translation units and affect configuration files.
Since the error log (e.g., a compilation error) contains the paths to the files with errors, \sys localizes all affected files and asks the LLM to repair them with the corresponding contextual metadata.

\section{Implementation and Experimental Setup} \label{sec:setup}

% \definecolor{darkgreen}{RGB}{0,100,0}

This section describes the implementation and experimental setup.

\begin{table}[t]
    \centering \setlength{\tabcolsep}{1.2pt} 
        \caption{Statistics of our benchmark dataset, intermediate results (\# Units and Unit size), testcases, and the Rust code translated by \sys. \textcolor{black}{Note that although \texttt{time} contains 20,436 Lines of Code (LoC), only 2,007 LoC are active code and others are dead code, e.g., \texttt{.h} files that are never included.}} %\todo{@momoko, please add \# of test cases for each program, and whether they are built-in or created by us.}
\label{target}
    \footnotesize
    \begin{tabular}{lrrrrrrr}
    \toprule
    \textbf{Program} & \multicolumn{6}{c}{\textbf{C Stat.}} & \multicolumn{1}{c}{\textbf{Rust Stat.}}\\
    \cmidrule(lr){2-7} \cmidrule(lr){8-8}
     & LoC  & \# Logic & \# Units & Unit Size  & \# Test  & Type & LoC \\
     &   & Blocks &  &  \\
    % \textbf{Program} & \multicolumn{4}{c|}{\textbf{C}} & \multicolumn{4}{c}{\textbf{Rust}}\\
    %  & C LoC & \# Files & Avg LoC & \multicolumn{1}{c|}{\# Blocks} & Rust LoC & \multicolumn{1}{c}{Passed} & \# ptrs & \# uses\\
    % & \# Files & Avg LoC
    \midrule
bst & \textcolor{black}{100} & 10 & 4 & 143.00 ± 0.00 & \textcolor{black}{4} & Built-in & \textcolor{black}{130} \\
buffer & \textcolor{black}{527} & 59 & 8 & 255.50 ± 5.50 & 17 & Built-in & \textcolor{black}{741} \\
ht & \textcolor{black}{717} & 95 & 36 & 97.22 ± 50.53 & 6 & Built-in & \textcolor{black}{632} \\
urlparser & \textcolor{black}{469} & 38 & 8 & 270.00 ± 170.00 & 41 & Built-in & \textcolor{black}{497} \\
SipHash~\cite{SipHash} & \textcolor{black}{3,189} & 63 & 23 & 406.12 ± 362.83 & 20 & Created & 671 \\
sds~\cite{sds} & \textcolor{black}{1,943} & 117 & 15 & 245.50 ± 173.08 & 24 & Built-in & \textcolor{black}{1,662} \\
rgba & \textcolor{black}{497} & 33 & 8 & 241.50 ± 159.50 & 5 & Built-in & \textcolor{black}{747} \\
quadtree & \textcolor{black}{513} & 42 & 8 & 188.00 ± 84.00 & 4 & Built-in & 515 \\
qsort & \textcolor{black}{69} & 14 & 8 & 59.50 ± 29.50 & \textcolor{black}{20} & \textcolor{black}{Provided} & \textcolor{black}{164} \\
mark-sweep~\cite{mark-sweep} & \textcolor{black}{244} & 25 & 4 & 266.00 ± 0.00 & \textcolor{black}{5} & Built-in & 381 \\
avl & \textcolor{black}{181} & 13 & 4 & 278.00 ± 0.00 & 2 & Built-in & \textcolor{black}{297} \\
which & 3,530 & 442 & 30 & 343.78 ± 340.85 & 20 & Created & \textcolor{black}{2,234} \\
Tinyhttpd~\cite{Tinyhttpd} & \textcolor{black}{418} & 39 & 8 & 234.50 ± 196.50 & \textcolor{black}{3} & Built-in & \textcolor{black}{354} \\
tiny-AES-c~\cite{tiny-AES-c} & \textcolor{black}{719} & 65 & 8 & 419.50 ± 119.50 & \textcolor{black}{3} & Built-in & \textcolor{black}{738} \\
time & \textcolor{black}{20,436} & 812 & 80 & 90.05 ± 101.74 & \textcolor{black}{3} & Built-in & \textcolor{black}{1,913} \\
c4~\cite{c4} & \textcolor{black}{502} & 12 & 4 & 325.00 ± 0.00 & \textcolor{black}{2} & Built-in & 990 \\
yank~\cite{yank} & \textcolor{black}{465} & 45 & 4 & 496.00 ± 0.00 & \textcolor{black}{22} & \textcolor{black}{Provided} & \textcolor{black}{578} \\
su-exec~\cite{su-exec} & \textcolor{black}{130} & 12 & 4 & 114.00 ± 0.00 & \textcolor{black}{20} & Created & \textcolor{black}{213} \\
\tiny{FastestWebsiteEver} ~\cite{FastestWebsiteEver} & \textcolor{black}{188} & 21 & 4 & 162.00 ± 0.00 & \textcolor{black}{5} & Built-in & 129 \\
mcrcon~\cite{mcrcon} & \textcolor{black}{717} & 50 & 4 & 664.00 ± 0.00 & \textcolor{black}{30} & Created & \textcolor{black}{589} \\
\textcolor{black}{zopfli} & \textcolor{black}{3,545} & \textcolor{black}{464} & \textcolor{black}{44} & \textcolor{black}{271.53 ±  227.24} & \textcolor{black}{20} & \textcolor{black}{Created} & \textcolor{black}{4,131} \\
%cflow & 12256 & 1023 & 56 &&   \\
    \bottomrule
    \end{tabular}
\end{table}

\begin{table*}[!t]
\footnotesize
%\color{blue} 
%\setlength{\tabcolsep}{3.0pt} %\setlength\columnsep{2pt}
\setlength{\tabcolsep}{1.7pt}
\caption{[RQ1] Comparison of \sys with baseslines on safety-related metrics (uLoC ($\downarrow$): Unsafe Lines of Code, \#RawDecl ($\downarrow$): Number of Raw Pointer Declarations, and \#RawDeref ($\downarrow$): Number of Raw Pointer Dereferences). The numbers under \#RawDecl and \#RawDeref have the format of N (M), where N is the output from the raw pointer counter of C2SaferRust, and M is the output from the counter of Crown.  Note that ``-'' means that the code cannot be complied and thus C2SaferRust fails to count the number of raw pointer declarations or references. The light blue highlighting (\colorbox{blue!15}{}) denotes the lowest value.}
\label{tab:rq1}
\begin{tabular}{lccccccccccccccc}
\toprule
\multirow{2}{*}{\textbf{Program}} & \multicolumn{3}{c}{\textbf{C2Rust}} & \multicolumn{3}{c}{\textbf{Crown}} & \multicolumn{3}{c}{\textbf{Laertes}} & \multicolumn{3}{c}{\textbf{C2SaferRust}} & \multicolumn{3}{c}{\textbf{\sys}} \\
 \cmidrule(lr){2-4} \cmidrule(lr){5-7} \cmidrule(lr){8-10} \cmidrule(lr){11-13} \cmidrule(lr){14-16} 
& uLoC & \#RawDecl & \#RawDeref & \multicolumn{1}{c}{uLoC} & \multicolumn{1}{c}{\#RawDecl} & \multicolumn{1}{c}{\#RawDeref} & \multicolumn{1}{c}{uLoC}  & \multicolumn{1}{c}{\#RawDecl} & \multicolumn{1}{c}{\#RawDeref} & \multicolumn{1}{c}{uLoC} &  \multicolumn{1}{c}{\#RawDecl} &  \multicolumn{1}{c}{\#RawDeref} & \multicolumn{1}{c}{uLoC}& \multicolumn{1}{c}{\#RawDecl} & \multicolumn{1}{c}{\#RawDeref} \\
\midrule
bst & 96 & 10(9) & 29 (26)& 61 & 4(4) & 7(7) & 106 & 10(10) & 29(29) & 68 & 9(9) & 26(26) & \cellcolor{blue!15}0 & \cellcolor{blue!15}0(0) & \cellcolor{blue!15}0(0) \\
buffer & 1,006 & 61(39) & 82(56) & 780 & 19(57) & 19(70) & 816 & (58) & (70) & 410 & 41(38) & 47(36) & \cellcolor{blue!15}0 & \cellcolor{blue!15}0(0) & \cellcolor{blue!15}0(0) \\
ht & 1,117 & 77(32) & 126(31) & 178 & (19) & (43) & 1,017 & 77(77) & 120(87) & 773 & 55(28) & 85(22) & \cellcolor{blue!15}0 & \cellcolor{blue!15}0(0) & \cellcolor{blue!15}0(0) \\
urlparser & 1,035 & 79(22) & 60(42) & 1,967 & (78) & (42) & 1,924 & 74(79) & 2(60) & 633 & 81(20) & 51(1) & \cellcolor{blue!15}0 & \cellcolor{blue!15}0(0) & \cellcolor{blue!15}0(0) \\
SipHash & 598 & 12(0) & 70(0) & 598 & 12(0) & 70(0) & 1,318 & 12(12) & 70(0) & 516 & 12(0) & 55(0) & \cellcolor{blue!15}0 & \cellcolor{blue!15}0(0) & \cellcolor{blue!15}0(0) \\
sds & 2,271 & 142(89) & 209(22) & 2,276 & (74) & (20) & 2,270 & (130) & (86) & - & - & - & \cellcolor{blue!15}0 & \cellcolor{blue!15}0(0) & \cellcolor{blue!15}0(0) \\
rgba & 994 & 21(21) & 78(64) & 604 & 16(21) & 66(64) & 573 & 21(21) & 78(64) & 263 & 5(5) & 8(6) & \cellcolor{blue!15}0 & \cellcolor{blue!15}0(0) & \cellcolor{blue!15}0(0) \\
quadtree & 905 & 47(38) & 209(104) & 1,664 & 33(47) & 114(156) & 905 & 47(38) & 209(104) & 387 & 35(35) & 106(73) & \cellcolor{blue!15}0 & \cellcolor{blue!15}0(0) & \cellcolor{blue!15}0(0) \\
qsort & 156 & 4(11) & 10(8) & 156 & (11) & (8) & 131 & 2(13) & 6(8) & 19 & 1(1) & 6(0) & \cellcolor{blue!15}0 & \cellcolor{blue!15}0(0) & \cellcolor{blue!15}0(0) \\
mark-sweep & 278 & 27(24) & 64(49) & 278 & (21) & (38) & 278 & 27(24) & 64(49) & 130 & 20(20) & 44(42) & \cellcolor{blue!15}0 & \cellcolor{blue!15}0(0) & \cellcolor{blue!15}0(0) \\
avl & 159 & 17(16) & 83(75) & 84 & 5(13) & 8(50) & 215 & 17(17) & 83(79) & 159 & 16(16) & 80(74) & \cellcolor{blue!15}0 & \cellcolor{blue!15}0(0) & \cellcolor{blue!15}0(0) \\
which & 1,666 & (68) & (81) & 1,669 & (29) & (47) & 1,666 & - & - & - & - & - & \cellcolor{blue!15}24 & \cellcolor{blue!15}0(0) & \cellcolor{blue!15}0(0) \\
Tinyhttpd & 1,742 & 13 (3) & 16 (0) & 1,742 & (3) & (0) & 1,180 & 12(12) & 16(7) & 1,520 & 10(1) & 13(0) & \cellcolor{blue!15}0 & \cellcolor{blue!15}0(0) & \cellcolor{blue!15}0(0) \\
tiny-AES-c & 1,522 & 36(36) & 126(100) & 1,517 & 30(30) & 47(20) & 1,176 & 36(36) & 126(100) & 791 & 14(14) & 35(10) & \cellcolor{blue!15}0 & \cellcolor{blue!15}0(0) & \cellcolor{blue!15}0(0) \\
time & 1,313 & 59(49) & 171(97) & 1,313 & (21) & (86) & 1,234 & (55) & (97) & 941 & 45(33) & 140(66) & \cellcolor{blue!15}121 & \cellcolor{blue!15}0(0) & \cellcolor{blue!15}0(0) \\
c4 & 1,685 & 41(9) & 319(9) & 1,681 & (9) & (9) & 1,690 & (34) & (60) & 1,727 & 48(45) & 214(69) & \cellcolor{blue!15}0 & \cellcolor{blue!15}0(0) & \cellcolor{blue!15}0(0) \\
yank & 770 & 14(9) & 47(8) & 764 & (6) & (8) & 729 & (12) & (11) & 709 & 11(6) & 39(0) & \cellcolor{blue!15}26 & \cellcolor{blue!15}0(0) & \cellcolor{blue!15}0(0) \\
su-exec & 145 & 9(0) & 15(0) & 145 & (0) & (0) & 119 & 9(9) & 15(8) & 134 & 9(0) & 15(0) & \cellcolor{blue!15}0 & \cellcolor{blue!15}0(0) & \cellcolor{blue!15}0(0) \\
FastestWebsiteEver & 203 & 5(1) & 9(1) & 203 & (1) & (1) & 173 & (5) & (7) & 197 & 6(2) & 8(2) & \cellcolor{blue!15}0 & \cellcolor{blue!15}0(0) & \cellcolor{blue!15}0(0) \\
mcrcon & 12,981 & 21(15) & 33(17) & 12,981 & (5) & (14) & 1,325 & 21(21) & 33(24) & 8,648 & 15(15) & 28(24) & \cellcolor{blue!15}0 & \cellcolor{blue!15}0(0) & \cellcolor{blue!15}0(0) \\
\midrule
$\Delta$ C2Rust  & - & - & - & \multicolumn{1}{c}{+0.1\%} & \multicolumn{1}{c}{(-8.6\%)} & \multicolumn{1}{c}{(-13.5\%)} & \multicolumn{1}{c}{-38.5\%} & \multicolumn{1}{c}{(+48.9\%)} & \multicolumn{1}{c}{(+30.5\%)} & \multicolumn{1}{c}{-43.8\%} &\multicolumn{1}{c}{-37.7\%} &\multicolumn{1}{c}{-43.1\%} &\multicolumn{1}{c}{-99.4\%} &\multicolumn{1}{c}{-100.0\%} &\multicolumn{1}{c}{-100.0\%} \\
\bottomrule
\end{tabular}
\end{table*}

\vspace{0.05in}
\noindent{\bf Implementation.} \hspace{0.05in}
Our implementation contains 38,341 Lines of Python Code. The default LLM is based on Claude 3.7 Sonnet (claude-3-7-sonnet-20250219)~\cite{3_7_sonnet}.
%Our open-source implementation~\cite{add} contains 38,341 Lines of Python Code. The default LLM is based on Claude 3.7 Sonnet (claude-3-7-sonnet-20250219)~\cite{3_7_sonnet}.
\sys's parser uses \texttt{libclang}~\cite{libclang}, a C interface to the Clang compiler and provides access to Abstract Syntax Tree (AST) of the target C code. 
 
Response guidelines specify the format for LLM responses.
To enable autonomous corrections, we define an API with three modes, requiring the LLM to select one in each response:  
(i) \textbf{read\_data}: reads the content of specified files. If necessary, specific line ranges can be designated, (ii) \textbf{modify\_data} : modifies or deletes line ranges in specified files, and (iii) \textbf{execute\_command} : executes shell scripts or other commands.
% \begin{itemize}
%     \item \textbf{read\_data} : Reads the content of specified files. If necessary, specific line ranges can be designated.
%     \item \textbf{modify\_data} : Modifies line ranges in specified files. 
%     %The affected lines are explicitly defined, and an option ensures all intended features are fully preserved.
%     \item \textbf{execute\_command} : Executes (multiple) commands.
% \end{itemize}

\textcolor{black}{
Our evaluation procedure includes two shell scripts to automate building and test execution in a feedback loop.  
These scripts consist of:  
(1) a script for building the translated Rust program, and 
%(2) a script for building the original C program,  
(2) a script for running tests and validating semantic equivalence.
%(4) a build script that performs all of these tasks. % (\texttt{run\_all.sh}).
%\sys runs test cases by calling the Rust programs from C programs using the Foreign Function Interface (FFI).  
The shell script for verifying semantic equivalence in (2) records the results from the C golden implementation in advance and compares them with the actual execution results of the Rust program.
This approach is adopted due to two considerations. 
%First, since the translated Rust program is untrusted and serves as the verification target, internal assertion checks in the translated Rust code cannot be relied upon. 
First, executing tests directly within the C program would necessitate FFI calls to invoke Rust functions from C, resulting in C-ish type representations for the translated Rust functions. Second, this external verification approach enables LLMs to provide feedback during the repair cycle based on a formalized response format. The shell script is specifically designed to display test numbers, execution results, and expected values in a structured manner, allowing LLMs to address multiple test cases sequentially rather than attempting to fix multiple failed tests simultaneously.
}
% Additionally, each function execution trace is saved to a designated path, ensuring LLMs can access and analyze the execution flow for informed repair decisions.

\textcolor{black}{
To implement the semantic equivalence described above, for programs that are not standalone, we rewrite the developer-written test cases even if they exist.
Specifically, the method involves three steps: (i) generates a C test program that calls the (library) function and displays the result. (ii) Translates both the library function and the test program to Rust. (iii) Uses the shell script to compare the outputs of both versions.
}
\textcolor{black}{
In \sys, tests are executed not by the LLM itself, but by the  software program of \sys. Therefore, print funcatations are needed to inform the LLM of the execution results.
}

% We use the APIs of Claude 3.7 Sonnet (claude-3-5-sonnet-20241022), since several studies~\cite{} have evaluated it as being skilled in agentic coding. 
% All code for preprocessing, translation, and repair is implemented in Python to facilitate Claude API calls. The parser uses \texttt{libclang}, which is a C interface to the Clang compiler and provides an access to Abstract Syntax Tree (AST) of the target C code.

% %Also, we attempted to use Llama 3-70b-instruct~\cite{llama3}, which has a context window size of 8,000 tokens. The input size exceeded the context window limit even for the smallest program (\texttt{bst}). Additionally, the responses did not follow the intended format. Therefore, we adopted Claude 3.7 Sonnet.

% While \sys basically works automatically, i.e., has the LLM handle the translation and repair processes, we perform manual operations in two scenarios.
% (1) The modifications to file structure and linking for the FFI calls are performed manually between the 1st repair and 2nd repair loops.
% % While \sys completes the functional repair check for programs up to 3,000 lines, \sys is also able to successfully compile the 5,400-line \texttt{lil} and 12,256-line \texttt{clow}.
% % For \texttt{clow}, the total number of repair iterations needed to successfully compile is [number] times.
% (2) When running test cases and the translated code contains an error triggering infinite loops, we manually fix the program.

\vspace{0.05in}
\noindent{\bf Setup.} \hspace{0.05in}
 Our experiment is conducted on an Azure Standard\_L8as\_v3 virtual machine with 512 GB of memory running Ubunutu 22.04 LTS.
 In our experiment, the default segment size parameter is set as 400 lines. Note that since segmentation is done without breaking logical boundaries, not all unit sizes are exactly 400 lines.
To reduce probabilistic behavior, the temperature~\cite{temperature} is set to 0, and \texttt{max\_tokens} parameter~\cite{temperature}, which indicates the maximum output tokens, is set to 8,192.

% Regarding test cases

\vspace{0.05in}
\noindent{\bf Dataset.} \hspace{0.05in} 
\begin{comment}
We curated a dataset of C programs as shown in Table~\ref{target}.
Our curation methodology is as follows. First, we selected the top C programs on GitHub ranked by stars, which has 100--2,000 lines of code. We target programs within this specific range because in state-of-the-art works~\cite{pan2024lost, yang2024vert, eniser2024towards, li2024translating}, even those with just 100 lines often fail to compile. After mining the repositories, our filtering criteria includes: programs with more than 500 stars, non-interactive execution, available test cases or documentation, and platform independence.
Second, we select 10 programs from prior works: 8 programs from Crown~\cite{zhang2023ownership} or Laertes~\cite{emre2023aliasing} (under 2,000 lines and non-interactive) and 2 GNU package programs (up to 3,000 lines) studied in~\cite{hong2024don}.
In total, our dataset consists of 20 programs: 10 from GitHub repositories (\texttt{SipHash}, \texttt{sds}, \texttt{mark-sweep}, \texttt{Tinyhttpd}, \texttt{FastestWebsiteEver}, \texttt{c4}, \texttt{yank}, \texttt{su-exec}, \texttt{tiny-ASE-c}, and \texttt{mcrcon}) and 10 from prior works.
\end{comment}
Table~\ref{target} shows our curated dataset of 21 C programs: 10 from top GitHub repositories (100--2,000 lines, 500+ stars, non-interactive) and 11 from prior works~\cite{zhang2023ownership, emre2023aliasing, hong2024don, shetty2024syzygy}.
The GitHub programs are: \texttt{SipHash}, \texttt{sds}, \texttt{mark-sweep}, \texttt{Tinyhttpd}, \texttt{FastestWebsiteEver}, \texttt{c4}, \texttt{yank}, \texttt{su-exec}, \texttt{tiny-ASE-c}, and \texttt{mcrcon}.
 %\autoref{target} summarizes the statistics for the target programs.
``\# Logic Blocks'' denotes the number of parsed elements, 
``\# Units'' the number of translation units, ``\# Unit Size'' the average size of the units, and 
``\# Test'' the number of test cases (based on the number of assertions).
% \textcolor{black}{
% We also classify the test cases based on their ``Type'', i.e., how we obtained them. 
% }
\textcolor{black}{
Test ``Type'' indicates the source: ``built-in'' (from original repository), ``created'' (manually developed with expected values/behaviors), or ``provided'' (from third-party repositories\footnote{E.g., \texttt{qsort} tests: \url{https://github.com/bradtraversy/traversy-js-challenges/blob/main/09-sorting-algorithms/11-quick-sort-implementation/quick-sort-test.js}; \texttt{yank} tests: \url{https://formulae.brew.sh/formula/yank}}).
}
\begin{comment}
Specifically, there are three sources: 
%Type represents the test type. Tests are classified to better reflect their nature:
\begin{icompact}
\item ``built-in'': We obtain the test cases from the original repository containing the source code.
\item ``created'': We create the test cases ourselves with the expected values and behaviors. 
\item ``provided'': We obtain the test cases from a third-party repository.  For example, the test cases of \texttt{qsort} are obtained from this third-party repo\footnote{\url{https://github.com/bradtraversy/traversy-js-challenges/blob/main/09-sorting-algorithms/
11-quick-sort-implementation/quick-sort-test.js}} and those of \texttt{yank} are from this repo\footnote{\url{https://formulae.brew.sh/formula/yank}}.
\end{icompact}
\end{comment}
%
% those whose expected values or behaviors are newly created by us will be termed `created.' They are in the original repository containing the source code are classified as `built-in.' They are not developed from scratch and come from third-party sources will be termed `provided.'
% %, i.e., whether the tests are built-in or created manually. 
% We use the provided sources if they are available; otherwise, we manually create test cases from scratch.  
% %To assure the reliability of test cases, we also consider coverage perspective, and ensure that test cases achieve at least 80 \% line coverage for all the programs.
%\textcolor{darkgreen}{When the provided sources are limited, we add additional created tests considering line coverage.}
\textcolor{black}{
Note that when verifying non-standalone Rust programs using C test cases, rewriting is necessary regardless of whether LLMs are used, so that the translated Rust program can be built and tested. In this situation, we have two choices: performing rewriting like Laertes, or excluding such programs from the target and testing only command line tools instead, as C2SaferRust does. We adopt the former approach to enable comparison with previous studies.
}

\vspace{0.05in}
\noindent{\bf Baselines.} \hspace{0.05in} In the evaluation, we adopt the following four baselines as a comparison with \sys. 

\begin{icompact}
\item C2Rust~\cite{c2rust}. \hspace{0.05in} C2Rust is a rule-based approach developed by Galois and Immunant for C-to-Rust translation. 
\item Crown~\cite{zhang2023ownership}. \hspace{0.05in} Crown is a rule-based ownership analysis tool for unsafe Rust code with the functionality of refactoring. 
\item C2SaferRust~\cite{c2saferrust}. \hspace{0.05in} C2SaferRust is a tool combining the strengths of both rule- and LLM-based approaches. It first adopts C2Rust for conversion and then utilizes LLMs for refinement. 
\item Laertes~\cite{emre2023aliasing}. \hspace{0.05in}  Laertes is a rule-based translation tool focusing on raw pointers in translated Rust code. 
\end{icompact}

\textcolor{black}{
Additionally, Syzygy~\cite{shetty2024syzygy} performs C-to-Rust translation using LLMs with formal equivalence verification methods. Although Syzygy did not publicly release their translation code itself, the translated Rust code for their target program (zopfli) is available. Therefore, we also conduct additional comparisons with Syzygy for zopfli.
}

\vspace{0.05in}
\noindent{\bf Evaluation Metrics.} \hspace{0.05in} We compare \sys with baselines based on two metrics, safety and semantic equivalence. 

%\begin{icompact}
%\item Safety.  \hspace{0.05in} 
First, the safety metric adopts three sub-metrics: (i) UnSafe Line of Code (uLoC), which measures the lines of code inside \texttt{unsafe} blocks and functions, (ii) Number of raw pointer declarations (\#RawDecl), i.e., the frequency of \texttt{*mut T} and \texttt{*const T} raw pointer type declarations, and (iii) Number of raw pointer dereferences (\#RawDeref), i.e., the count of operations where raw pointers are dereferenced. 
We use counters from both C2SaferRust (counts mutable and immutable pointers in compilable code) and Crown (counts only mutable pointers) for comprehensive comparison.

\textcolor{black}{
The semantic equivalence metric adopts three sub-metrics: (i) number of errors during the compilation stage, (ii) test case pass rate and (iii) ratio of C macros remaining in the translated code. Specifically, we measure the number of Rust variables which are originally C macro variables, Rust macro functions, and macros used for conditional compilation.
%\end{icompact}
}
\newcommand{\twounsafemean}{604.67}
\newcommand{\twouptrdeclmean}{866.5}
\newcommand{\twouptrrefmean}{768.5}
 %\newcommand{\twouptrrefperc}{}

% Crown values
\newcommand{\crownunsafemean}{780.00}
\newcommand{\crownptrdeclmean}{433.5}
\newcommand{\crownptrrefmean}{434.5}

% Laertes values
\newcommand{\laertesunsafemean}{532.52 }
\newcommand{\laertesptrdeclmean}{244.23 }
\newcommand{\laertesptrrefmean}{313.9}

% C2SaferRust values
\newcommand{\safunsafemean}{532.52 }
\newcommand{\safptrdeclmean}{244.23 }
\newcommand{\safptrrefmean}{313.9}

%%%%%%%%%% Fixed area %%%%%%%%%%
\newcommand{\crownunsafeperc}{-0.1\%\xspace}
\newcommand{\crownptrdeclperc}{8.6\%\xspace}
\newcommand{\crownptrrefperc}{13.5\%\xspace}

\newcommand{\laertesunsafeperc}{38.5\%\xspace}
\newcommand{\laertesptrdeclperc}{-48.9\%\xspace}
\newcommand{\laertesptrrefperc}{-30.5\%\xspace}

\newcommand{\safunsafeperc}{43.8\%\xspace}
\newcommand{\safptrdeclperc}{47.3\%\xspace}
\newcommand{\safptrrefperc}{45.7\%\xspace}
%%%%%%%%%%%%%%%%%%%%%%%%%%%%%%

% w/o Refinement values
\newcommand{\worefunsafemean}{3.23 }
\newcommand{\worefunsafeperc}{54.4\%\xspace}
\newcommand{\worefptrdeclmean}{0 }
\newcommand{\worefptrdeclperc}{30.8\%\xspace}
\newcommand{\worefptrrefmean}{0 }
\newcommand{\worefptrrefperc}{16.8\%\xspace}

% Self-Refinement values
% \newcommand{\selfrefunsafemean}{3.15 }
% \newcommand{\selfrefunsafeperc}{51.2\% }
% \newcommand{\selfrefptrdeclmean}{9.45 }
% \newcommand{\selfrefptrdeclperc}{48.4\% }
% \newcommand{\selfrefptrrefmean}{}
% \newcommand{\selfrefptrrefperc}{}

% Refinement values
\newcommand{\withrefunsafemean}{2.00 }
\newcommand{\withrefunsafeperc}{98.87\%\xspace}
\newcommand{\withrefptrdeclmean}{0 }
\newcommand{\withrefptrdeclperc}{0 \%\xspace}
\newcommand{\withrefptrrefmean}{0 }
\newcommand{\withrefptrrefperc}{0 \%\xspace}

\section{Evaluation} \label{sec:evaluation}
Our evaluation aims to address the following research questions:
\begin{icompact}
    \item \textbf{RQ1} [Safety]: How is \sys effective compare to existing translation approaches in terms of safety?
    \item \textbf{RQ2} [Semantics]: How does \sys compare with prior works in terms of semantic equivalence?

%    Can \sys scale its translation functionality with increasing LoC?
    \item \textbf{RQ3} [Performance+Cost]: What is the performance and cost overhead of \sys? 
    \item \textbf{RQ4} [Ablation Study]: How does each strategy of \sys contribute to the translation success?

\end{icompact}

\subsection{RQ1: Memory Safety}
\textcolor{black}{
%\todo{Describe Table~\ref{tab:rq1} here when it is ready.}
Table~\ref{tab:rq1} presents the result of safety metrics for each program. The bottom row ($\Delta$ C2Rust) shows the average change rate from C2Rust.
\sys shows significant improvements over existing methods.
Compared with C2Rust, on average, it reduces unsafe LoC from 1,550 to 8.14 (a 99.4 \% reduction), and raw pointer declarations and dereferences are significantly reduced across all programs.
This already outperforms Crown, which achieves \crownunsafeperc, \crownptrdeclperc, \crownptrrefperc reductions, Laertes, which manages \laertesunsafeperc, \laertesptrdeclperc, and \laertesptrrefperc reductions, and C2SaferRust, which achieves \safunsafeperc, 37.7 \%, 43.1 \% reductions respectively.
We observed that in Crown, Laertes, and C2SaferRust, the metrics increased rather than decreased for several programs.
\sys demonstrates a reduction in all metrics across every program, and even when unsafe lines are exist, it is limited to just 8.14 lines on average.
}

\textcolor{black}{
We also measured unsafe LoC for the moment before the LLM self-refines the code in stage (ii). We then observed that  the unsafe LoC is decreased in 11.8\% on average (standard deviation 93.9\%) at the moment when the LLM quality-refines its code. 
For 7 programs (\texttt{which}, \texttt{ht}, \texttt{sds}, \texttt{time}, \texttt{FastestWebsiteEver}, \texttt{mcrcon} and \texttt{su-exec}), the uLoC decreases. For example, the uLoC of \texttt{which} decreased from 843 to 53, and that of \texttt{FastestWebsiteEver} fell from 97 to 0 - all showing reduction rates exceeding 90\%.
In \sys, although we instruct the LLM to generate safe code from the very beginning of the translation process, some unsafe operations still appear in the initial output.
However, these unsafe operations are largely eliminated through the quality-refinement process.
}

% On the contrary, as exceptions, the uLoC of \texttt{TinyHttpd} has increased from 0 to 10. This is due to the adjustment of interface types to accommodate the C test case calls. The unsafe statements that remain in several programs are concentrated around the functions at the boundary between C and Rust code because these functions use C-native types.
% \sys does not directly include the unsafe count in the self-evaluation (as described in \autoref{self_check}) because it considers cases where unsafe operations are necessary.

% \autoref{tab:improvements} shows the number of unsafe LoC, raw pointer declarations, and uses for \sys, C2Rust and two previous studies for each program. The table also shows the metrics for both the initially translated code (w/o Refinement) right after providing the translation instructions in the \sys framework.  % and the subsequently improved code after LLM self-refinement (with Self-Refinement)

The safety improvements of \sys over C2Rust, Laertes, Crown, and C2SaferRust stem from four concrete advances: (1) precise C-to-Rust type mapping that preserves safety guarantees, (2) flexible replacement of unsafe external functions with safe alternatives, (3) careful management of global variables to prevent memory violations, and (4) strategic selection of Rust standard libraries that maximize safety while maintaining functionality.
% \begin{itemize}
% \item Diverse Type Mapping
% \item Flexible replacement of unsafe external function calls
% \item Handling of global variables
% \item Selection of appropriate standard libraries
% \end{itemize}

\textcolor{black}{In comparison with Syzygy, their publicly available zopfli code has no unsafe or raw pointers, compiles without errors, and passes all test cases. 
Regarding macros, \sys  retains 68 \% macros, while Syzygy does not handle macro transformations because they use the unifdef tool to expand conditional macros.
While they show comparable results on some metrics, the authors acknowledge several limitations that require manual intervention, as discussed in their challenges and future work section. Specifically, struct definitions and their field relationships must be manually translated due to the need for global program context. Additionally, when validation tests fail, their approach requires manual edits to the translated codebase. The authors recognize these limitations and identify developing more systematic LLM-guided repair techniques as future work. By contrast, \sys distinguishes itself by generating the resulting Rust code through LLM-driven automated translation and repair.}

\begin{figure}[!t]
\begin{lstlisting}[language=C, basicstyle=\ttfamily\footnotesize,caption=A snippet of C code with memory safety issues in urlparser~\cite{emre2023aliasing}, label={lst:case-1}]
  char *pathname = malloc(sizeof(char));
  // ...
  sscanf(tmp_path, "%[^?|^#]", pathname);// buf overflow
  // ...
  char *tmp_url = strdup(URL);
  char *protocol = url_get_protocol(tmp_url);
  data->protocol = protocol;
  // ...
  free(tmp_url);
\end{lstlisting}
\begin{lstlisting}[language=C, label={lst:case-1-rust}, basicstyle=\ttfamily\footnotesize,caption=The safe translation of Listing~\ref{lst:case-1} with \sys]
  fn new(url: &str) -> Option<Self> {
    let parsed_url = Url::parse(url).ok()?;
    // Safe pathname extraction
    let pathname = parsed_url.path().split('?').next()?.to_string();
    // Safe protocol extraction
    let protocol = parsed_url.scheme().to_string();
    
    Some(Self { // All fields have independent ownership
        pathname,
        protocol, // ... Other fields
    })
  }
\end{lstlisting} %\vspace{-0.3in}
\end{figure}

\begin{table}[!t]
\footnotesize \setlength{\tabcolsep}{2.1pt} %\footnotesize 
    \caption{[RQ2] Comparison of semantics-related metrics (\#Errors: total number of compilation errors, \%Tests: average percentage of passed test cases, and \%Marco: average percentage of correctly translated macros. Values in parentheses indicate standard deviation).} \label{tab:rq2} \vspace{-0.1in}
\begin{tabular}{lccccc}
\toprule
\multirow{1}{*}{\textbf{Metric}} & \textbf{C2Rust} & \textbf{Crown} & \textbf{Laertes} & \textbf{C2SaferRust}  & \textbf{\sys} \\
\midrule
\#Errors ($\downarrow$) & 9 & 253 & 13 & 9 & 0 \\
\%Tests ($\uparrow$) & 80 ($\pm$ 30) \% & 35 ($\pm$ 47) \% & 80 ($\pm$ 30) \% & 80 ($\pm$ 30) \% & 100 ($\pm$ 0) \% \\
\%Marco ($\uparrow$) & 0 ($\pm$ 0) \% &   0 ($\pm$ 0) \% &  0 ($\pm$ 0) \% &  0 ($\pm$ 0) \% & \textcolor{black}{85.2($\pm$ 18.6)} \% \\
\bottomrule
\end{tabular} \vspace{-0.1in}
\end{table}

% \#Errors & 9 (9) & 253 & 33 & 9 (543) & 0 (1,818) \\

\begin{table}[!t]
%\color{blue}
\footnotesize \setlength{\tabcolsep}{5.5pt}
    \caption{[RQ2] Breakdown of semantics-related metrics for SmartC2Rust (\#Errors: Number of intermediate errors, Passed: Whether test cases are passed, and \%Marco: Percentage of marcos preserved).} \label{tab:rq2_smart}
\begin{tabular}{lccccccc}
\toprule
\multirow{2}{*}{\textbf{Program}} & \multicolumn{4}{c}{\textbf{\# Intermediate Errors.}} & \textbf{Passed} & \textbf{\%Macro} \\
\cmidrule(lr){2-5}
 & \multicolumn{2}{c}{\textbf{Compile (\%)}} & \multicolumn{2}{c}{\textbf{Semantic  (\%)}} & \textbf{} & \textbf{} \\
\midrule
bst & 0 & (0.00\%) & 0 & (0.00\%) & \ding{51} & 100.0\% \\
buffer & 99 & (0.80\%) & 24 & (0.20\%) & \ding{51} & 75.0\% \\
ht & 64 & (0.82\%) & 14 & (0.18\%) & \ding{51} & 100.0\% \\
urlparser & 48 & (0.23\%) & 158 & (0.77\%) & \ding{51} & 100.0\% \\
SipHash & 2 & (0.20\%) & 8 & (0.80\%) & \ding{51} & 71.4\% \\
sds & 78 & (0.69\%) & 35 & (0.31\%) & \ding{51} & 79.2\% \\
rgba & 16 & (0.80\%) & 4 & (0.20\%) & \ding{51} & 100.0\% \\
quadtree & 55 & (0.90\%) & 6 & (0.10\%) & \ding{51} & 100.0\% \\
qsort & 6 & (1.00\%) & 0 & (0.00\%) & \ding{51} & - \\
mark-sweep & 2 & (0.40\%) & 3 & (0.60\%) & \ding{51} & 100.0\% \\
avl & 1 & (0.07\%) & 13 & (0.93\%) & \ding{51} & - \\
which & 79 & (0.68\%) & 37 & (0.32\%) & \ding{51} & 33.3\% \\
Tinyhttpd & 0 & (0.00\%) & 4 & (1.00\%) & \ding{51} & 100.0\% \\
tiny-AES-c & 53 & (0.90\%) & 6 & (0.10\%) & \ding{51} & 61.5\% \\
time & 133 & (0.25\%) & 399 & (0.75\%) & \ding{51} & 61.0\% \\
c4 & 73 & (0.53\%) & 65 & (0.47\%) & \ding{51} & 100.0\% \\
yank & 101 & (0.89\%) & 12 & (0.11\%) & \ding{51} & 91.7\% \\
su-exec & 31 & (0.91\%) & 3 & (0.09\%) & \ding{51} & - \\
FastestWebsiteEver & 3 & (0.38\%) & 5 & (0.62\%) & \ding{51} & 100.0\% \\
mcrcon & 74 & (0.00\%) & 2 & (1.00\%) & \ding{51} & 90.9\% \\
zopfli & 122 & (0.78\%) & 34 & (0.21\%) & \ding{51} & 68.9\% \\
\midrule
Total &  1,040 & - &  832 & - & 20/20 & 85.2\% \\
\bottomrule
\end{tabular}
\end{table}

\subsubsection{A Case Study}
 Listing~\ref{lst:case-1} shows a case study of memory safety issues in existing translation works.  C2Rust still allocates just one byte for pathname and performs unbounded writes, preserving the buffer overflow risk. It also maintains the same pointer dependency pattern that leads to use-after-free issues when \texttt{tmp\_url} is freed but protocol continues to be referenced. 
 In contrast, Listing~\ref{lst:case-1-rust} shows the translation by \sys: \texttt{parsed\_url.path()} and \texttt{parsed\_url.scheme()} return string slices (\&str type) and the \texttt{to\_string()} method automatically allocates exactly the right amount of memory. The use-after-free issue is eliminated by creating independent string values with their own memory allocations.

\subsection{RQ2: Semantic Equivalence}
Table~\ref{tab:rq2} shows the result of semantic-related metrics compared with baselines for all programs.
All Rust code produced by \sys are error-free, both in terms of compilation and semantics, resulting in 0 compilation errors and 100 \% passing rate.
In contrast, baseline approaches have difficulty in compilation and running tests.
In Crown, 10 programs failed to compile, and 2 programs (\texttt{sds}, \texttt{which}) did not compile with any baseline approaches. C2SaferRust encountered compilation errors and fixed them during rewriting unsafe statements using LLMs, but it is basically relied on the output of C2Rust and does not fix programs that initially fail to compile with C2Rust.
%Additionally, for running tests with rule-based methods, manual patches are prepared in advance as shown in ~\cite{}, so we.
For macros, since all of the baseline methods simply expand them, there is no equivalent of them in the translated output.
On the contrary, in \sys, on average, \textcolor{black}{85.2} \% of the original C macros remain in the transformed Rust code.

%\todo{Add texts when the table is ready. }

\subsubsection{Fixing Compilation Errors}
% \vspace{0.05in}
% \noindent{\bf Fixing Compilation Errors.}  \hspace{0.05in} 

\begin{table} 
\footnotesize
%\color{blue} 
\caption{[RQ2] Breakdown of encountered compilation errors during iterative refinement} \label{tab:breakdown}
%\todo{@Momoko, please expand others into details. }
\begin{tabular}{lc}
\toprule
{\bf Error Type (Rust's Error Codes Index)} & {\bf Percentage}\\
\midrule
E0308: Type mismatches & 19.13\%\\
E0425: Use of undeclared names & 17.44 \%\\
E0609: Access to nonexistent fields & 7.47 \%\\
E0432: Unresolved imports & 7.21 \%\\
E0433: Problems with importing named elements & 6.67\% \\
% E0599: Methods or associated functions not being found & 8.9\% \\
% E0433: Problems with importing named elements & 5.9\% \\
Others* &  \\
\multicolumn{2}{l}{* E0599 (5.60\%), E0502 (4.36\%), E0616 (3.56\%), E0597 (3.11\%), E0061 (3.02\%),  } \\
\multicolumn{2}{l}{ E0277 (3.02\%), E0428 (2.67\%), E0015 (2.31\%), E0382 (1.87\%), E0106 (1.33\%),  } \\
\multicolumn{2}{l}{ E0252 (1.33\%), E0659 (1.25\%), E0119 (1.07\%), E0499 (0.89\%), E0761 (0.89\%), } \\
\multicolumn{2}{l}{ E0282 (0.71\%), E0596 (0.71\%), E0560 (0.62\%), E0716 (0.62\%), E0107 (0.53\%), } \\
\multicolumn{2}{l}{ E0603 (0.36\%), E0605 (0.27\%), E0505 (0.27\%), E9 (0.27\%), E0614 (0.18\%), } \\
\multicolumn{2}{l}{ E0412 (0.18\%), E0506 (0.18\%), E0583 (0.18\%), E0515 (0.18\%), E0658 (0.18\%), } \\
\multicolumn{2}{l}{ E0133 (0.18\%), E0507 (0.18\%) } \\
% \multicolumn{2}{l}{* E0277 (4.46\%), E0502 (2.34\%), E0432 (1.99\%), E0106 (1.65\%), E0609 (1.44\%), } \\
% \multicolumn{2}{l}{E0499 (1.17\%), E0061 (1.10\%), E0515 (1.03\%), E0412 (1.03\%), E0505 (0.96\%), } \\
% \multicolumn{2}{l}{ E0616 (0.89\%), E0560 (0.76\%), E0503 (0.62\%), E0761 (0.55\%), E0119 (0.55\%),} \\
% \multicolumn{2}{l}{ E0107 (0.55\%), E0133 (0.48\%), E0606 (0.34\%), E0317 (0.34\%), E0255 (0.27\%), } \\
% \multicolumn{2}{l}{ E0506 (0.27\%), E0753 (0.27\%), E0596 (0.27\%), E0614 (0.14\%), E0593 (0.14\%), } \\
% \multicolumn{2}{l}{E0716 (0.14\%), E0382 (0.14\%), E0384 (0.14\%)} \\
\bottomrule
\end{tabular}
\end{table}

% \multicolumn{2}{l}{* E0277 (4.46\%), E0502 (2.34\%), E0432 (1.99\%), E0106 (1.65\%), E0609 (1.44\%), } \\
% \multicolumn{2}{l}{E0499 (1.17\%), E0061 (1.10\%), E0515 (1.03\%), E0412 (1.03\%), E0505 (0.96\%), } \\
% \multicolumn{2}{l}{E0616 (0.89\%), E0560 (0.76\%), E0503 (0.62\%), E0761 (0.55\%), E0119 (0.55\%), } \\
% \multicolumn{2}{l}{E0107 (0.55\%), E0133 (0.48\%), E0606 (0.34\%), E0317 (0.34\%), E0255 (0.27\%), }  \\
% \multicolumn{2}{l}{E0506 (0.27\%), E0753 (0.27\%), E0596 (0.27\%), E0614 (0.14\%), E0593 (0.14\%), } \\
% \multicolumn{2}{l}{E0716 (0.14\%), E0382 (0.14\%), E0384 (0.14\%) }\\

\textcolor{black}{
While \sys ultimately generates error-free Rust code, it resolved a total of 1,040 compile errors and 832 semantic errors during the iterative repair loop.
Table~\ref{tab:rq2_smart} presents the number of fixed errors for each program. % during \sys's iterative refinement for each program. 
While the prevalence of compile errors versus semantic errors varies by program, compile errors are on average 1.2 times more common because modifications made to address semantic issues can sometimes introduce new compilation errors.
}

Table~\ref{tab:breakdown} shows a breakdown of the type of encountered and fixed compilation errors. The errors are categorized according to Rust's error codes index~\cite{errorcode}.
\textcolor{black}{
The vast majority of the errors are related to dependencies.
 We observed that LLMs tend to initially follow instructions less often even when dependency summary information is included in the instruction. 
 In fact, on average, 52.13\% of all errors are resolved in the very first iteration, which means  the number of errors decreases significantly after the first iteration.
 }
This is likely because the first translation instructions contain many translation rules and response guidelines, making it difficult to follow each rule individually.
 %\autoref{down} describes the relationship between the number of compilation errors and iteration counts in the 1st repair process. 
 %It shows that 
Since \sys consistently provide dependency information within the iterative loop, the LLM begins to focus on dependency information and utilizes them once the errors occurred.
As a result, it prevents repeated compilation fails.

% \vspace{0.05in}
% \noindent{\bf Fixing Compilation Errors.}  \hspace{0.05in} 

\subsubsection{Fixing Semantics-related Errors}

Semantics-related errors vary depending on each program, such as incorrect borrowing syntax or improper parsing methods, which cause tests to fail. 
\sys has the LLM response with the reasoning behind the correction method, and we observed that
LLMs methodically address each test case step by step to implement the necessary corrections.
By repeating such iterative corrections, \sys eventually achieves test pass rate of 100 \% for all the programs.

%E0425 is the most prevalent at 26.8\%, followed by E0599 at 13.4\%, E0308 at 13.2\%, E0277 at 9.2 \%, and E0433 at 6 \%. E0425 indicates the use of undeclared names, E0599 represents methods or associated functions not being found, E0308 shows type mismatches, E0277 relates to unsatisfied trait bounds, and E0433 reflects problems with importing named elements.

\subsubsection{Macro Preservation} %Equality}
 %\subsubsection{Qualitative Analysis on Macros used in Conditional Compilation}
 Table~\ref{tab:rq2_smart} also shows the percentage of macros that are correctly translated.
 We handled a total of 1,061 C macros.
Out of these, 50 are macro functions, and 164 are macros used for conditional compilation.
The two GNU packages, \texttt{which} and \texttt{time} contain a large number of conditional macros (88 and 47 respectively) due to their emphasis on portability.
%Note that all methods relying on rule-based translation (Crown, Laertes and C2SaferRust) expand macro expressions before translation, so there is no Rust equivalent of the C macros in the translated output.

C macro variables are correctly replaced with Rust const variables. 
Of the conditional macros, 26\% are replaced with Rust Cargo macros. %44 are replaced with Rust Cargo macros.
Then, we observed that 29\% of conditional macros have no corresponding equivalents in Rust. %48 macros have no corresponding equivalents in Rust.
The representative is include guards of C macros. % 48 macros
 For header guards, these guards can typically be identified heuristically through naming conventions (ending in \texttt{\_H})~\cite{header}, as there is no strict definition to distinguish them from other conditional directives using \texttt{\#ifndef}. 
 %While this naming convention is generally reliable, there are exceptions like \texttt{HAVE\_CONFIG\_H}. 
 The LLM identified correctly include guards and macros purely used for conditional compilation. 
However, for several cases, we observed inappropriate translation.
For example, the LLM interpreted \texttt{HAVE\_CONFIG\_H} as a Cargo feature flag. Since \texttt{HAVE\_CONFIG\_H} is specifically tied to C's Autoconf/Automake system~\cite{autoconf, automake} for checking the presence of \texttt{config.h}, and Rust's build system takes a fundamentally different approach to configuration management, the flag seems unnecessary.
We also observed that C macro functions are replaced with either Rust macro functions, ordinary functions or Rust standard library functions.

 % The inability to use Rust's cfg attribute significantly hampers code maintainability and flexibility. Without this feature, developers cannot implement conditional compilation within the same source file, forcing them to create duplicate code for different platforms or features. This not only increases code complexity but also reduces portability across different systems.

 \begin{figure}[!t] 
\begin{lstlisting}[language=C, caption={A snippet of tiny-AES-c with conditional compilation (only \sys translates correctly).}, label={lst:conditional_c}, basicstyle=\ttfamily\footnotesize, label={lst:case-2}]
  #if (defined(CBC) && (CBC == 1)) || (defined(CTR) && (CTR == 1))
  void AES_init_ctx_iv(struct AES_ctx* ctx, const uint8_t* key, const uint8_t* iv){
    KeyExpansion(ctx->RoundKey, key);
    memcpy (ctx->Iv, iv, AES_BLOCKLEN);
  }
  #endif
\end{lstlisting}
%\end{figure}
%
%\begin{figure}[t] 
\begin{lstlisting}[language=C, caption={Correct translations of Listing~\ref{lst:case-2} by \sys.}, label={lst:conditional_rust}, basicstyle=\ttfamily\footnotesize]
  #[cfg(any(feature = "cbc", feature = "ctr"))]
  pub fn init_ctx_iv(ctx: &mut AES_ctx, key: &[u8], iv: &[u8]) {
    key_expansion(&mut ctx.round_key, key);
    ctx.iv.copy_from_slice(iv);
  }
\end{lstlisting}
\end{figure}

\subsubsection{A Case Study}  

Listing~\ref{lst:case-2} shows another case study of semantic equivalence issues, especially about macro transformation in existing translations. All existing works, including C2Rust, Crown, Laertes, and C2SaferRust, adopts a preprocessor for processing macros, which removed the \texttt{\#if} condition at Line 1 of Listing~\ref{lst:case-2}, violating the semantics of the code.  Instead, \autoref{lst:conditional_rust} shows the correct translation by \sys, which consolidates the preprocessor conditions using Rust's \texttt{\#[cfg]} attributes. \sys enables this conditioned translation because it recognizes macros (\texttt{CBC} and \texttt{CBC}) during pre-processing of \sys scheme, and the LLM determines to handle these macros as Cargo features and set in \texttt{Cargo.toml} at the very beginning of the translation. During the translation of each unit, \sys then applies these feature flags to the appropriate code sections.

\subsection{RQ3: Overhead}

We evaluate two types of overhead: (1) performance in time (seconds), iterations, and number of tokens and chats, and (2) cost in US dollars. 
\autoref{tab:performance} shows the performance and cost for each program.

%\subsubsection{Performance}
\noindent{\bf Performance.} \hspace{0.05in} 
\textcolor{black}{
As shown in the table, 
%for example, converting \texttt{sds} (with about 1,800 lines of code) already 
on average, \sys requires a total of 187 user and LLM chats, 10,456 K input tokens, 116 K output tokens, and 2,637 seconds to translate.
}
%One key feature of \sys is that the LLM selects the mode type and iteratively refines the code interactively, therefore . 
This is because \sys employs an interactive approach where the LLM selects the mode type and iteratively refines the code, which inherently demands a moderate number of chat interactions and processing time.
The input tokens are particularly long because \sys sends back the file that the LLM requests to modify when making corrections, along with line number information to ensure that the LLM correctly identifies the range to be revised.

% In \sys, when making corrections, it sends back the file that the LLM requests to modify along with line number information to ensure that the LLM correctly identifies the sections to be revised. As a result, the input tokens are particularly long.

%\subsubsection{Cost} 
\noindent{\bf Cost.} \hspace{0.05in} 
Claude 3.7 Sonnet's pricing is based on token usage. The cost for processing input tokens is 3 US dollars per million tokens (MTok), while the cost for output tokens is 15 US dollars per million tokens (MTok)~\cite{pricing}.
Therefore, as shown in ~\autoref{tab:performance}, the target program size relates with the cost, but it also correlates with the number of chat requests/responses and iteration count.
%as the input and output tokens increase, the cost increases. 
%For example, \texttt{sds}, which has frequent interaction (378 chats) with the LLM, takes 58 USD. Overall, these cost is reasonable to translate.
\textcolor{black}{
On average, the translation requires 33 USD. It is reasonable considering that users get high-quality Rust code without any unsafe statements.
}

\begin{table}[!t]
    \centering
    %\color{blue} 
\setlength{\tabcolsep}{2pt}
    \caption{[RQ3] Performance overhead and cost}
    \label{tab:performance}
    %\todo{@Momoko, please match the programs with Tables 2, 3 and 4.  Thanks! The list is different at this moment. }
    \footnotesize % temporary
    \begin{tabular}{lrrrrrrrr}
    \toprule
   \multirow{2}{*}{\textbf{Program}} & \multirow{2}{*}{\textbf{\#Chats}}  & \multicolumn{2}{c}{\textbf{\#Iterations}}  & \multirow{2}{*}{\textbf{Time} [s]} & \multicolumn{2}{c}{\textbf{\#Token [K] }}  & \textbf{Cost} \\ % & \multicolumn{1}{c}{\textbf{\#Error}} & \multicolumn{2}{c}{\textbf{Pass Rate}} \\
\cmidrule(lr){3-4} \cmidrule(lr){6-7}
 & & \textbf{Comp.} & \textbf{Semant.} &  & In & Out  & [USD] \\ %&  \textbf{} & \textbf{} & \textbf{} & \textbf{(\%)} \\ % \#CE % \#SE
    \midrule
bst & 19 & 1 & 1 & 149.19 & 182.19 & 17.75 & 0.81 \\
buffer & 103 & 6 & 3 & 1,106.81 & 3,899.12 & 78.51 & 12.87 \\
ht & 265 & 140 & 4 & 1,984.37 & 8,597.70 & 147.71 & 28.01 \\
urlparser & 112 & 7 & 11 & 1,263 & 4,454.94 & 62.98 & 14.31 \\
SipHash & 177 & 50 & 0 & 1,623.81 & 6,679.59 & 184.40 & 22.80 \\
sds & 243 & 15 & 7 & 2,789.64 & 21,361.38 & 91.06 & 65.45 \\
rgba & 127 & 15 & 6 & 1,402.35 & 4,655.77 & 74.98 & 15.09 \\
quadtree & 447 & 17 & 93 & 10,235.62 & 41,768.83 & 213.45 & 128.51 \\
qsort & 20 & 5 & 0 & 2,380.37 & 183.81 & 13.93 & 0.76 \\
mark-sweep & 47 & 1 & 3 & 325.12 & 1,046.67 & 25.82 & 3.53 \\
avl & 113 & 6 & 12 & 1,250.81 & 5,503.83 & 70.22 & 17.56 \\
which & 472 & 141 & 19 & 4,014.03 & 24,989.98 & 306.60 & 79.57 \\
Tinyhttpd & 62 & 5 & 4 & 463.66 & 1,112.14 & 36.80 & 3.89 \\
tiny-AES-c & 118 & 15 & 0 & 3,702.23 & 8,308.98 & 136.60 & 26.98 \\
time & 697 & 448 & 89 & 10,621.62 & 35,032.16 & 358.88 & 110.48 \\
c4 & 100 & 3 & 0 & 1,178.09 & 6,122.78 & 103.65 & 19.92 \\
yank & 98 & 16 & 0 & 1,229.34 & 9,757.17 & 72.83 & 30.36 \\
su-exec & 36 & 6 & 0 & 261.93 & 909.18 & 18.96 & 3.01 \\
FastestWebsiteEver & 26 & 3 & 0 & 154.95 & 400.36 & 16.15 & 1.44 \\
mcrcon & 74 & 6 & 1 & 613.12 & 2,737.58 & 51.62 & 8.99 \\
zopfli & 590 & 42 & 23 & 8,642.67 & 31,886.72 & 358.88 & 101.04 \\
\midrule
%Average & 167.8  & 45.3 & 12.65 & 2,337.50 & 9,385.21  &  104.14    & 33.12   \\
Average & 187.90 & 45.14 & 13.14 & 2,637.75 & 10,456.71 & 116.27  & 33.12   \\

    \bottomrule
    \end{tabular}

\end{table}

\subsection{RQ4: Ablation Study}

\newcommand{\randoma}{28.5\%\xspace}
\newcommand{\summarya}{32.3\%\xspace}
\newcommand{\moda}{12.4\%\xspace}
\newcommand{\flowa}{23.3\%\xspace}

\begin{table}[t]
    \caption{[RQ4] Ablation study}
    \label{tab:definition-order} \setlength{\tabcolsep}{6.5pt} 
    \centering \footnotesize
    \begin{tabular}{lrr}
        \toprule
        \multirow{1}{*}{\textbf{Strategy}} & \textbf{Compilation Success \%}  & \textbf{Test Case Pass \%} \\
     %   & \textbf{} & \textbf{}  \\
        \midrule
        Random Order & \randoma & - \\
        w/o Contextual Summary & \summarya & - \\
        Fixed File Modification  & \moda & - \\
        w/o Flow comparison & - & \flowa \\
        \midrule
        \sys & \textbf{100.0 \%} & \textbf{100.0 \%} \\
        \bottomrule
    \end{tabular}

\end{table}

We conducted an ablation study to understand different strategy adopted by \sys.
We use three C programs, \texttt{urlparser}, \texttt{sds}, and \texttt{which}.  Our evaluation is based on the following two metrics: (i) compilation success percentage (\%), i.e., the average percentage of LoC that successfully compiled out of the total LoC, and (ii) test case pass rate, i.e., the average proportion of successful test cases out of all test cases.

%We conducted the following ablation studies on \texttt{urlparser}, \texttt{sds}, and \texttt{which}.
%\autoref{tab:definition-order} compares the success rate of each repair loop  with cases where each strategy was not applied. The 1st repair success rate refers to the average percentage of LoC that successfully compiled out of the total LoC. The 2nd repair success rate indicates the average proportion of successful test cases out of all test cases.

%\begin{itemize}

\noindent{\bf Definition-Order Translation.} \hspace{0.05in} 
In this study, we make the order of the files translation as random.
% Similar to our proposal, this baseline method has no memory between conversions, but memory is retained during the repair process. 
\autoref{tab:definition-order} shows that in the random order, the success rate is \randoma. % when the order of definitions is not sorted
%The primary compilation errors that LLMs cannot solve are as follows.
When an LLM calls a function without knowing its definition, we observed that it takes one of three approaches. The LLM either guesses the implementation based on the function name and context, replaces it with \texttt{unimplemented()!}, or assumes it exists in an external package.
Such bold code generation ultimately leads to a problem where errors are repeated and compilation fails to converge.

\noindent{\bf Contextual Dependency Summarization.} \hspace{0.05in} 
%\autoref{tab:definition-order} also reveals the significant impact of maintaining contextual summaries of dependencies between translation units. 
 In this study, we omitted these contextual summaries from our translation process and 
\autoref{tab:definition-order} shows that the success rate dropped to just \summarya.
The Rust compiler primarily reveals dependency-related errors (E0308, E0599, E0425, E0432) and field access violations (E0609, E0616). 
%These errors stem from insufficient understanding of module boundaries and visibility rules.
Without contextual dependency feedback, translators are difficult to identify and resolve interface inconsistencies within the limited number of compilation attempts.
%, such as functions expecting \texttt{&mut unit1::SdsString} but receiving \texttt{&mut RefMut<'_, module1_h::unit0::SdsString>} instead.

\noindent{\bf Mode-based API.} \hspace{0.05in} 
In this study, we let  
 the conversion to be performed sequentially in segments without modifying the other units/files. \autoref{tab:definition-order} shows that the compilation success rate is \moda.
%The following is an example where an LLM made modifications to a previously translated unit.
The most critical case where compilation becomes impossible is when the LLM needs external crates by modifying \texttt{Cargo.toml}. The crates are for example, \texttt{lazy\_static}~\cite{lazy_static} and \texttt{once\_cell}~\cite{once_cell}. 
If flexible modifications for these other units/files are not permitted, the errors will permanently remain unresolved.

\noindent{\bf Flow-based Localization in Semantic Repair.} \hspace{0.05in} 
 In this study, we omitted function execution flow feedback.  
 \autoref{tab:definition-order} shows that
%without function execution flow feedback, 
 the average semantic equivalence success rate is 23\%.
LLMs with poor localization abilities repeat the same code fixes in modify\_data mode when handling compile/semantic errors, or misdiagnose problems as permission/lock file issues, proposing cleanup and rebuild commands in execute\_command mode instead of addressing actual code defects.

\noindent{\bf Other LLMs beyond Claude.}  \hspace{0.05in} We also tested the programs using GPT-4o (gpt-4o-2024-08-06)~\cite{gpt_4o} and Gemini Pro 1.5 (gemini-pro)~\cite{gemini_pro}.
%, and Llama 3-70b-instruct (meta/meta-llama-3-70b-instruct)~\cite{llama3}.
GPT-4o achieved a compilation success rate of 34.3 \%, while Gemini Pro 1.5 had a lower rate of 12.4 \%. GPT-4o typically demonstrates consistent error patterns, whereas Gemini Pro 1.5 frequently produces inappropriately formatted responses. This performance discrepancy can be attributed to the differing levels of coding proficiency among LLMs, where Claude is superior to its competitors~\cite{aider}.
\textcolor{black}{
Open-source smaller LLMs usually have even more constrained context window sizes, making them prone to exceeding the context limit. We would expect to reduce \sys's context window and increase the iterations to work with smaller open-source models. We will leave this as our future work. 
}

\newcommand{\calcDiff}[2]{%
  \pgfmathparse{(#2-#1)/#2*100}%
  \pgfmathprintnumber[precision=1]{\pgfmathresult}\%
}

% Crown
\renewcommand{\crownunsafeperc}{\calcDiff{\crownunsafemean}{\twounsafemean}}
\renewcommand{\crownptrdeclperc}{\calcDiff{\crownptrdeclmean}{\twouptrdeclmean}}
\renewcommand{\crownptrrefperc}{\calcDiff{\crownptrrefmean}{\twouptrrefmean}}

% Laertes
\renewcommand{\laertesunsafeperc}{\calcDiff{\laertesunsafemean}{\twounsafemean}}
\renewcommand{\laertesptrdeclperc}{\calcDiff{\laertesptrdeclmean}{\twouptrdeclmean}}
\renewcommand{\laertesptrrefperc}{\calcDiff{\laertesptrrefmean}{\twouptrrefmean}}

% w/o Refinement
\renewcommand{\worefunsafeperc}{\calcDiff{\worefunsafemean}{\twounsafemean}}
\renewcommand{\worefptrdeclperc}{\calcDiff{\worefptrdeclmean}{\twouptrdeclmean}}
\renewcommand{\worefptrrefperc}{\calcDiff{\worefptrrefmean}{\twouptrrefmean}}

% Refinement
\renewcommand{\withrefunsafeperc}{\calcDiff{\withrefunsafemean}{\twounsafemean}}
\renewcommand{\withrefptrdeclperc}{\calcDiff{\withrefptrdeclmean}{\twouptrdeclmean}}
\renewcommand{\withrefptrrefperc}{\calcDiff{\withrefptrrefmean}{\twouptrrefmean}}

\section{Discussions and Limitations}  \label{sec:threat}

%The primary threat to the validity of our translation scheme is that the prompt's translation instructions might not ensure the quality of the generated Rust code.
%For example, instructions to declare all items (structures, enums, functions, constants, etc.) with pub (public) so they can be imported from other modules may not be desirable from the security reason.
%To address this, a solution might be to provide a refinement step where the visibility requirements of each element are analyzed and public access is restricted only to components that genuinely need it. 

We discuss a few commonly issues and \sys's limitations. 

%\vspace{0.05in}
\vspace{0.05in}
\noindent{\bf Scalability.}  \hspace{0.05in} 
We also tested a large program with 12,256 lines of code~\cite{cflow}. It compiled successfully, but the semantic repair encountered an infinite loop undetectable by the iterative process. Note that the compilation repair required 4,213 chats, costing 1,451 USD and taking 27 hours.
After manual fixes and LLM iterations, 24 \% of the test cases passed, but we observed that it remains challenging to push forward because the LLM is simplifying and rewriting the code just to make the test cases pass. As a future work, it is important to design instructions that do not oversimplify during semantic repair.

\vspace{0.05in}
\noindent{\bf Testing of different macro combinations.}  \hspace{0.05in} 
 Regarding conditional compilation with macros, \sys does not cover all the combination of macro definitions. For instance, with n macro variables \texttt{MACRO1}, \texttt{MACRO2}, ..., \texttt{MACRO\_n}, there are numerous possible states depending on whether each macro is defined and what values they hold. 
 %To claim complete translation validation, we would need to run the repair loop for all these combinations. 
 However, this limitation exists in the original C code as well—even with available macros, developers rarely test all possible conditional compilation paths. Nevertheless, testing compile and semantic repair under key macro condition combinations would generate more sustainable Rust code.

\vspace{0.05in}
\noindent{\bf LLM's awareness of the target C program during training.}  \hspace{0.05in} 
 \sys instructs LLMs to translate only within the provided C code range, even if they recognize the entire program. This restriction maintains our segmented translation approach but may prevent optimizations that could benefit from the LLM's whole-program understanding. Future work should explore methods to harness LLMs' broader knowledge while preserving consistency across code segments.
 
\vspace{0.05in}
\noindent{\textcolor{black}{\noindent\bf Adaptation for languages other than C.}  \hspace{0.05in}} 
\textcolor{black}{Although substantial engineering work remains, the core principles of our methodology—code segmentation and feedback-driven iterative loops—are language-agnostic and can be extended to other programming languages to tackle similar fundamental challenges, including LLM context window constraints, syntactic error resolution, and semantic correctness verification.}

\section{Related Work} \label{sec:relate}

%\todo{@Momoko, Can you add related works in C2SaferRust?  The number of citations should be at least between 30 and 40.  Thank you!}

\noindent{\bf LLM-based C-to-Rust translation.} \hspace{0.05in} 
%\emph{\textbf{LLM-base C-to-Rust Translation:}}
%Verified
Pan et al.~\cite{pan2024lost} analyzed translation patterns of LLMs across multiple language pairs, including C to Rust. They demonstrated that compilation errors persist even in attempts to translate small C programs.
%even when targeting small C programs, compilation errors occur, making conversion impossible.
% While this work mentions the potential of LLMs to generate safer Rust code compared to rule-based approaches, it does not address the appropriate prompts or specific mechanisms for generate safe Rust code.
% It also demonstrates that even when targeting small programs, compilation errors occur, making conversion impossible.
%There is little work on LLM-based C-to-Rust translation.
Yang et al.~\cite{yang2024vert} presented a formal verification framework for validating semantic equivalence following the translation of target C programs via LLMs.
%Towards
Eniser et al.~\cite{eniser2024towards} also utilized LLMs for translation and applied fuzzing techniques to verify equivalence.
However, both approaches can convert only C programs consisting of several hundred lines of code into compilable Rust code: Yang et al.~\cite{yang2024vert} fails to compile code with 158 lines and Eniser et al.~\cite{eniser2024towards} can handle only code between 150 and 597 lines.
Li et al.~\cite{li2024translating} investigated translation patterns among human participants, LLM-based translators, and rule-based translators. They found the state-of-the-art LLM-based translators produce idiomatic safe Rust snippets yet fail to generate compilable code due to type inconsistencies when merging independently translated functions. 
%This identified challenge remains unresolved and is designated for future investigation.
% but not compilable programs. They stated the reason the compilation fails is as when functions are translated in separate components, the LLM often produces inconsistent and conflicting type definitions that cannot be reconciled when attempting to merge them back into a single program. 
% Although they mentioned the observation, the challenge remains as future work.
% % lost in transaction

C2SaferRust~\cite{nitin2025c2saferrust} translated 7 real-world programs by using an LLM to eliminate unsafe statements created by C2Rust.
However, this approach exhibits two main limitations: it cannot translate non-function code elements such as structure definitions, nor can it substitute C FFI function calls with safe Rust equivalents. 
Consequently, in some cases, the translator output may have more unsafe lines of code or raw pointers compared to C2Rust.

% One of its limitations is that it does not translate code outside of functions, such as structure definitions.
% Additionally, C2SaferRust cannot replace C FFI function calls with their safe Rust equivalents.
% As a result, in some cases, the translator output may have more unsafe lines of code or raw pointers compared to C2Rust.

Hong et.al ~\cite{hong2025type} translated C function signatures into Rust using LLMs, but only fixed type compilation errors, excluding functions with large bodies from translation. Other compilation errors remain uncorrected, and the semantic equivalence is not resolved.

\textcolor{black}{
While Syzygy~\cite{shetty2024syzygy} also translated C code to Rust, it has limitations including lack of iterative repair process, translation of struct definitions, and support for macro translation. These functions are currently implemented manually in the translated Rust code, though the authors identify developing more systematic techniques for LLM-guided translation and repair as future work.
}

%  and multi-file programs, which are essential features for translating real-world C codebases

\vspace{0.05in}
\textcolor{black}{\noindent{\bf LLM-based translation for other language pairs.} 
Some LLM-based translators~\cite{ibrahimzada2024program, yang2024exploring, xueinterpretable, dearing2024lassi, yin2024rectifier, bhattarai2024enhancing} other than C to Rust have been proposed.
For example, Ibrahimzada~\cite{ibrahimzada2024program} suggested a method-level, partial translation approach from Python to Java code that takes into account the size limit of context windows.
%However, they did not address the challenges inherent in C to Rust translations in terms of memory safety.
}
\textcolor{black}{Zhang et al.~\cite{zhang2025scalable} converts Go code to Rust, which involves different language paradigms and safety considerations compared to C-to-Rust translation. 
}
 \textcolor{black}{
These concurrent works, while relevant to the broader domain of automated language translation, tackle distinct problems with different technical constraints and evaluation methodologies than \sys approach.
}

\begin{comment}
\paragraph{Other Language Translation.}
%\emph{\textbf{Other Language Translation:}}
% これを1番最初に参照すべきなのか
~\cite{ibrahimzada2024program}
\end{comment}

%
% c2rust
% https://github.com/immunant/c2rust
\vspace{0.05in}
\noindent{\bf Rule-based C-to-Rust translation.} \hspace{0.05in} 
%\emph{\textbf{Handcrafted conversion method:}}
A representative rule-based translation tool is the C2Rust Transpiler~\cite{c2rust}, which uses predefined rules to automatically translate C code into Rust. However, C2Rust generates code with excessive unsafe and unidiomatic Rust code. It struggles with pointer handling, external libraries, and performance optimization, requiring significant manual refactoring. 
Thus, several rule-based translation methods ~\cite{emre2021translating, hong2023concrat, emre2023aliasing, hong2024don} have been proposed to achieve better conversions beyond C2Rust, such as replacing unsafe raw pointers with safe references~\cite{emre2021translating, hong2023concrat, emre2023aliasing}, handling lock APIs~\cite{hong2023concrat}, and focusing on tuples and Option/Result types~\cite{hong2024don}.
Specific targets can be reliably transformed using the translation mapping, but the translated code still contains various unsafe. Moreover, context-dependent translation is not feasible.
% While rule-based approaches have the advantage of being able to handle large-scale code, LLM-based approaches have the potential to produce more natural and secure translations with less effort.

% ~\cite{hong2023concrat}, ~\cite{emre2023aliasing}
% ~\cite{hong2024don}
% ~\cite{zhang2023ownership}
\vspace{0.05in}
\noindent{\bf Program repair.} \hspace{0.05in} 
Many studies proposed LLM-based program repair schemes~\cite{deligiannis2023fixing, silva2023repairllama, zhang2024systematic, bouzenia2024repairagent, jimenez2024swebench}.
%\emph{\textbf{Compile Repair:}}
Particularly, Deligiannis et al.~\cite{deligiannis2023fixing} demonstrated an approach for fixing Rust compilation errors using formatted logs. 
%a method of using an LLM to fix Rust code by utilizing formatted compilation error logs.
While they have a similar iterative correction process, \sys distinctively supports multi-file modifications through LLM mode selection and incorporates additional data retrieval and command execution capabilities.
% The overall process of repeatedly fixing compilation errors remains the same, but our approach typically differs in that we have enabled the modification of multiple files based on the LLM's mode choices and also provided additional tools such as for reading data or executing commands.

\begin{comment}
% これはトピックとして必要？
\paragraph{Prompt Size and Performance.}
%\emph{\textbf{Prompt Size and Performance:}}
\end{comment}
\section{Conclusion} \label{sec:conc}
This paper proposes a method, called \sys, for translating C code into Rust using a large language model (LLM).
 Our key insight is an iterative refinement process in which LLMs progressively improve Rust code by incorporating feedback on semantic and memory safety issues until the code successfully passes test cases. We evaluated \sys against \textcolor{black}{21} programs up to approximately 3,000 lines of code and showed that \sys outperforms prior works, including C2Rust, Crown, Laertes, and C2SaferRust in terms of memory safety, i.e., reducing unsafe statements by \textcolor{black}{99.4}\%, and semantic equivalence, i.e., improving test case pass rates by 11.25\%. \sys also achieves \textcolor{black}{85.2}\% preservation of the semantic meaning of C macros.

\bibliographystyle{ACM-Reference-Format}
%\bibliography{sample-base}
\bibliography{bib.bib}


\begin{thebibliography}{60}

%%% the Web address.

\ifx \showCODEN    \undefined \def \showCODEN     #1{\unskip}     \fi
\ifx \showISBNx    \undefined \def \showISBNx     #1{\unskip}     \fi
\ifx \showISBNxiii \undefined \def \showISBNxiii  #1{\unskip}     \fi
\ifx \showISSN     \undefined \def \showISSN      #1{\unskip}     \fi
\ifx \showLCCN     \undefined \def \showLCCN      #1{\unskip}     \fi
\ifx \shownote     \undefined \def \shownote      #1{#1}          \fi
\ifx \showarticletitle \undefined \def \showarticletitle #1{#1}   \fi
\ifx \showURL      \undefined \def \showURL       {\relax}        \fi
% invisible to TeX
\providecommand\bibfield[2]{#2}
\providecommand\bibinfo[2]{#2}
\providecommand\natexlab[1]{#1}
\providecommand\showeprint[2][]{arXiv:#2}

\bibitem[add(2025)]%
        {add}
 \bibinfo{year}{2025}\natexlab{}.
\newblock \bibinfo{title}{smartC2Rust Demo server}.
\newblock \bibinfo{howpublished}{\url{https://github.com/momo-trip/SmartC2Rust/tree/main/demo}}.
\newblock


\bibitem[rep(2025)]%
        {repo}
 \bibinfo{year}{2025}\natexlab{}.
\newblock \bibinfo{title}{smartC2Rust repository}.
\newblock \bibinfo{howpublished}{\url{https://github.com/momo-trip/SmartC2Rust}}.
\newblock


\bibitem[AI(2025a)]%
        {gemini_pro}
\bibfield{author}{\bibinfo{person}{Google AI}.} \bibinfo{year}{2025}\natexlab{a}.
\newblock \bibinfo{title}{Gemini 2.0 Pro}.
\newblock \bibinfo{howpublished}{\url{https://deepmind.google/technologies/gemini/pro/}}.
\newblock


\bibitem[AI(2025b)]%
        {gpt_4o}
\bibfield{author}{\bibinfo{person}{Open AI}.} \bibinfo{year}{2025}\natexlab{b}.
\newblock \bibinfo{title}{Hello GPT-4o}.
\newblock \bibinfo{howpublished}{\url{https://openai.com/index/hello-gpt-4o/}}.
\newblock


\bibitem[aider(2025)]%
        {aider}
\bibfield{author}{\bibinfo{person}{aider}.} \bibinfo{year}{2025}\natexlab{}.
\newblock \bibinfo{title}{Aider LLM Leaderboards}.
\newblock \bibinfo{howpublished}{\url{https://aider.chat/docs/leaderboards/}}.
\newblock


\bibitem[Anthropic(2024)]%
        {window}
\bibfield{author}{\bibinfo{person}{Anthropic}.} \bibinfo{year}{2024}\natexlab{}.
\newblock \bibinfo{title}{Context window}.
\newblock \bibinfo{howpublished}{\url{https://support.anthropic.com/en/articles/7996848-how-large-is-claude-s-context-window }}.
\newblock


\bibitem[Anthropic(2025a)]%
        {pricing}
\bibfield{author}{\bibinfo{person}{Anthropic}.} \bibinfo{year}{2025}\natexlab{a}.
\newblock \bibinfo{title}{Anthropic API Pricing}.
\newblock \bibinfo{howpublished}{\url{https://www.anthropic.com/pricing\#anthropic-api}}.
\newblock


\bibitem[Anthropic(2025b)]%
        {3_7_sonnet}
\bibfield{author}{\bibinfo{person}{Anthropic}.} \bibinfo{year}{2025}\natexlab{b}.
\newblock \bibinfo{title}{Claude 3.7 Sonnet and Claude Code}.
\newblock \bibinfo{howpublished}{\url{https://www.anthropic.com/news/claude-3-7-sonnet}}.
\newblock


\bibitem[Anthropic(2025c)]%
        {temperature}
\bibfield{author}{\bibinfo{person}{Anthropic}.} \bibinfo{year}{2025}\natexlab{c}.
\newblock \bibinfo{title}{Strengthen guardrails}.
\newblock \bibinfo{howpublished}{\url{https://docs.anthropic.com/en/docs/test-and-evaluate/strengthen-guardrails/reduce-latency}}.
\newblock


\bibitem[Bhattarai et~al\mbox{.}(2024)]%
        {bhattarai2024enhancing}
\bibfield{author}{\bibinfo{person}{Manish Bhattarai}, \bibinfo{person}{Javier~E Santos}, \bibinfo{person}{Shawn Jones}, \bibinfo{person}{Ayan Biswas}, \bibinfo{person}{Boian Alexandrov}, {and} \bibinfo{person}{Daniel O'Malley}.} \bibinfo{year}{2024}\natexlab{}.
\newblock \showarticletitle{Enhancing Code Translation in Language Models with Few-Shot Learning via Retrieval-Augmented Generation}.
\newblock \bibinfo{journal}{\emph{arXiv preprint arXiv:2407.19619}} (\bibinfo{year}{2024}).
\newblock


\bibitem[Bouzenia et~al\mbox{.}(2024)]%
        {bouzenia2024repairagent}
\bibfield{author}{\bibinfo{person}{Islem Bouzenia}, \bibinfo{person}{Premkumar Devanbu}, {and} \bibinfo{person}{Michael Pradel}.} \bibinfo{year}{2024}\natexlab{}.
\newblock \showarticletitle{Repairagent: An autonomous, llm-based agent for program repair}.
\newblock \bibinfo{journal}{\emph{arXiv preprint arXiv:2403.17134}} (\bibinfo{year}{2024}).
\newblock


\bibitem[c-code style(2025)]%
        {header}
\bibfield{author}{\bibinfo{person}{c-code style}.} \bibinfo{year}{2025}\natexlab{}.
\newblock \bibinfo{title}{Recommended C style and coding rules}.
\newblock \bibinfo{howpublished}{\url{https://github.com/MaJerle/c-code-style/blob/main/README.md\#headersource-files}}.
\newblock


\bibitem[c4(2025)]%
        {c4}
\bibfield{author}{\bibinfo{person}{c4}.} \bibinfo{year}{2025}\natexlab{}.
\newblock \bibinfo{title}{c4}.
\newblock \bibinfo{howpublished}{\url{https://github.com/rswier/c4}}.
\newblock


\bibitem[cflow(2025)]%
        {cflow}
\bibfield{author}{\bibinfo{person}{cflow}.} \bibinfo{year}{2025}\natexlab{}.
\newblock \bibinfo{title}{cflow}.
\newblock \bibinfo{howpublished}{\url{https://www.gnu.org/software/cflow/}}.
\newblock


\bibitem[Chen et~al\mbox{.}(2023)]%
        {chen2023frugalgpt}
\bibfield{author}{\bibinfo{person}{Lingjiao Chen}, \bibinfo{person}{Matei Zaharia}, {and} \bibinfo{person}{James Zou}.} \bibinfo{year}{2023}\natexlab{}.
\newblock \showarticletitle{Frugalgpt: How to use large language models while reducing cost and improving performance}.
\newblock \bibinfo{journal}{\emph{arXiv preprint arXiv:2305.05176}} (\bibinfo{year}{2023}).
\newblock


\bibitem[clang 20.0.0git(2025)]%
        {libclang}
\bibfield{author}{\bibinfo{person}{clang 20.0.0git}.} \bibinfo{year}{2025}\natexlab{}.
\newblock \bibinfo{title}{libclang: C Interface to Clang}.
\newblock \bibinfo{howpublished}{\url{https://clang.llvm.org/doxygen/group__CINDEX.html}}.
\newblock


\bibitem[crates.io(2025a)]%
        {lazy_static}
\bibfield{author}{\bibinfo{person}{crates.io}.} \bibinfo{year}{2025}\natexlab{a}.
\newblock \bibinfo{title}{lazy\_static v1.5.0}.
\newblock \bibinfo{howpublished}{\url{https://crates.io/crates/lazy_static}}.
\newblock


\bibitem[crates.io(2025b)]%
        {once_cell}
\bibfield{author}{\bibinfo{person}{crates.io}.} \bibinfo{year}{2025}\natexlab{b}.
\newblock \bibinfo{title}{once\_cell v1.21.1}.
\newblock \bibinfo{howpublished}{\url{https://crates.io/crates/once_cell}}.
\newblock


\bibitem[Cybersecurity et~al\mbox{.}(2003)]%
        {cisa2023}
\bibfield{author}{\bibinfo{person}{United~States Cybersecurity}, \bibinfo{person}{Infrastructure~Security Agency}, \bibinfo{person}{United States National Security Agency United States Federal~Bureau of Investigation}, \bibinfo{person}{Australian Signals Directorate’s Australian Cyber~Security Centre}, \bibinfo{person}{Canadian~Centre for Cyber~Security}, \bibinfo{person}{United Kingdom National Cyber~Security Centre}, \bibinfo{person}{New Zealand National Cyber~Security Centre}, {and} \bibinfo{person}{Computer Emergency Response Team~New Zealand}.} \bibinfo{year}{2003}\natexlab{}.
\newblock \bibinfo{title}{The Case for Memory Safe Roadmaps: Why Both C-Suite Executives and Technical Experts Need to Take Memory Safe Coding Seriously}.
\newblock


\bibitem[Dearing et~al\mbox{.}(2024)]%
        {dearing2024lassi}
\bibfield{author}{\bibinfo{person}{Matthew~T Dearing}, \bibinfo{person}{Yiheng Tao}, \bibinfo{person}{Xingfu Wu}, \bibinfo{person}{Zhiling Lan}, {and} \bibinfo{person}{Valerie Taylor}.} \bibinfo{year}{2024}\natexlab{}.
\newblock \showarticletitle{LASSI: An LLM-based Automated Self-Correcting Pipeline for Translating Parallel Scientific Codes}.
\newblock \bibinfo{journal}{\emph{arXiv preprint arXiv:2407.01638}} (\bibinfo{year}{2024}).
\newblock


\bibitem[Deligiannis et~al\mbox{.}(2023)]%
        {deligiannis2023fixing}
\bibfield{author}{\bibinfo{person}{Pantazis Deligiannis}, \bibinfo{person}{Akash Lal}, \bibinfo{person}{Nikita Mehrotra}, {and} \bibinfo{person}{Aseem Rastogi}.} \bibinfo{year}{2023}\natexlab{}.
\newblock \showarticletitle{Fixing rust compilation errors using llms}.
\newblock \bibinfo{journal}{\emph{arXiv preprint arXiv:2308.05177}} (\bibinfo{year}{2023}).
\newblock


\bibitem[Docs.rs(2025)]%
        {std_url}
\bibfield{author}{\bibinfo{person}{Docs.rs}.} \bibinfo{year}{2025}\natexlab{}.
\newblock \bibinfo{title}{Crate url}.
\newblock \bibinfo{howpublished}{\url{https://docs.rs/url/latest/url/}}.
\newblock


\bibitem[Emre et~al\mbox{.}(2023)]%
        {emre2023aliasing}
\bibfield{author}{\bibinfo{person}{Mehmet Emre}, \bibinfo{person}{Peter Boyland}, \bibinfo{person}{Aesha Parekh}, \bibinfo{person}{Ryan Schroeder}, \bibinfo{person}{Kyle Dewey}, {and} \bibinfo{person}{Ben Hardekopf}.} \bibinfo{year}{2023}\natexlab{}.
\newblock \showarticletitle{Aliasing limits on translating c to safe rust}.
\newblock \bibinfo{journal}{\emph{Proceedings of the ACM on Programming Languages}} \bibinfo{volume}{7}, \bibinfo{number}{OOPSLA1} (\bibinfo{year}{2023}), \bibinfo{pages}{551--579}.
\newblock


\bibitem[Emre et~al\mbox{.}(2021)]%
        {emre2021translating}
\bibfield{author}{\bibinfo{person}{Mehmet Emre}, \bibinfo{person}{Ryan Schroeder}, \bibinfo{person}{Kyle Dewey}, {and} \bibinfo{person}{Ben Hardekopf}.} \bibinfo{year}{2021}\natexlab{}.
\newblock \showarticletitle{Translating C to safer Rust}.
\newblock \bibinfo{journal}{\emph{Proceedings of the ACM on Programming Languages}} \bibinfo{volume}{5}, \bibinfo{number}{OOPSLA} (\bibinfo{year}{2021}), \bibinfo{pages}{1--29}.
\newblock


\bibitem[Eniser et~al\mbox{.}(2024)]%
        {eniser2024towards}
\bibfield{author}{\bibinfo{person}{Hasan~Ferit Eniser}, \bibinfo{person}{Hanliang Zhang}, \bibinfo{person}{Cristina David}, \bibinfo{person}{Meng Wang}, \bibinfo{person}{Brandon Paulsen}, \bibinfo{person}{Joey Dodds}, {and} \bibinfo{person}{Daniel Kroening}.} \bibinfo{year}{2024}\natexlab{}.
\newblock \showarticletitle{Towards Translating Real-World Code with LLMs: A Study of Translating to Rust}.
\newblock \bibinfo{journal}{\emph{arXiv preprint arXiv:2405.11514}} (\bibinfo{year}{2024}).
\newblock


\bibitem[FastestWebsiteEver(2025)]%
        {FastestWebsiteEver}
\bibfield{author}{\bibinfo{person}{FastestWebsiteEver}.} \bibinfo{year}{2025}\natexlab{}.
\newblock \bibinfo{title}{FastestWebsiteEver}.
\newblock \bibinfo{howpublished}{\url{https://github.com/diracdeltas/FastestWebsiteEver}}.
\newblock


\bibitem[Hanley(2023)]%
        {horizon3.ai}
\bibfield{author}{\bibinfo{person}{Zach Hanley}.} \bibinfo{year}{2023}\natexlab{}.
\newblock \bibinfo{title}{Rust Won’t Save Us: An Analysis of 2023’s Known Exploited Vulnerabilities}.
\newblock
\urldef\tempurl%
\url{https://www.horizon3.ai/attack-research/attack-blogs/analysis-of-2023s-known-exploited-vulnerabilities/}
\showURL{%
\tempurl}


\bibitem[Hong and Ryu(2023)]%
        {hong2023concrat}
\bibfield{author}{\bibinfo{person}{Jaemin Hong} {and} \bibinfo{person}{Sukyoung Ryu}.} \bibinfo{year}{2023}\natexlab{}.
\newblock \showarticletitle{Concrat: An automatic C-to-Rust lock API translator for concurrent programs}. In \bibinfo{booktitle}{\emph{2023 IEEE/ACM 45th International Conference on Software Engineering (ICSE)}}. IEEE, \bibinfo{pages}{716--728}.
\newblock


\bibitem[Hong and Ryu(2024)]%
        {hong2024don}
\bibfield{author}{\bibinfo{person}{Jaemin Hong} {and} \bibinfo{person}{Sukyoung Ryu}.} \bibinfo{year}{2024}\natexlab{}.
\newblock \showarticletitle{Don’t Write, but Return: Replacing Output Parameters with Algebraic Data Types in C-to-Rust Translation}.
\newblock \bibinfo{journal}{\emph{Proceedings of the ACM on Programming Languages}} \bibinfo{volume}{8}, \bibinfo{number}{PLDI} (\bibinfo{year}{2024}), \bibinfo{pages}{716--740}.
\newblock


\bibitem[Hong and Ryu(2025)]%
        {hong2025type}
\bibfield{author}{\bibinfo{person}{Jaemin Hong} {and} \bibinfo{person}{Sukyoung Ryu}.} \bibinfo{year}{2025}\natexlab{}.
\newblock \showarticletitle{Type-migrating C-to-Rust translation using a large language model}.
\newblock \bibinfo{journal}{\emph{Empirical Software Engineering}} \bibinfo{volume}{30}, \bibinfo{number}{1} (\bibinfo{year}{2025}), \bibinfo{pages}{3}.
\newblock


\bibitem[Ibrahimzada(2024)]%
        {ibrahimzada2024program}
\bibfield{author}{\bibinfo{person}{Ali~Reza Ibrahimzada}.} \bibinfo{year}{2024}\natexlab{}.
\newblock \showarticletitle{Program Decomposition and Translation with Static Analysis}. In \bibinfo{booktitle}{\emph{Proceedings of the 2024 IEEE/ACM 46th International Conference on Software Engineering: Companion Proceedings}}. \bibinfo{pages}{453--455}.
\newblock


\bibitem[Immunant(2022)]%
        {c2rust}
\bibfield{author}{\bibinfo{person}{Immunant}.} \bibinfo{year}{2022}\natexlab{}.
\newblock \bibinfo{title}{C2Rust}.
\newblock \bibinfo{howpublished}{\url{https://github.com/immunant/c2rust}}.
\newblock


\bibitem[Jimenez et~al\mbox{.}(2024)]%
        {jimenez2024swebench}
\bibfield{author}{\bibinfo{person}{Carlos~E Jimenez}, \bibinfo{person}{John Yang}, \bibinfo{person}{Alexander Wettig}, \bibinfo{person}{Shunyu Yao}, \bibinfo{person}{Kexin Pei}, \bibinfo{person}{Ofir Press}, {and} \bibinfo{person}{Karthik~R Narasimhan}.} \bibinfo{year}{2024}\natexlab{}.
\newblock \showarticletitle{{SWE}-bench: Can Language Models Resolve Real-world Github Issues?}. In \bibinfo{booktitle}{\emph{The Twelfth International Conference on Learning Representations}}.
\newblock
\urldef\tempurl%
\url{https://openreview.net/forum?id=VTF8yNQM66}
\showURL{%
\tempurl}


\bibitem[Language(2025)]%
        {errorcode}
\bibfield{author}{\bibinfo{person}{Rust~Programming Language}.} \bibinfo{year}{2025}\natexlab{}.
\newblock \bibinfo{title}{Rust error codes index}.
\newblock \bibinfo{howpublished}{\url{https://doc.rust-lang.org/error_codes/error-index.html}}.
\newblock


\bibitem[Levy et~al\mbox{.}(2024)]%
        {levy2024same}
\bibfield{author}{\bibinfo{person}{Mosh Levy}, \bibinfo{person}{Alon Jacoby}, {and} \bibinfo{person}{Yoav Goldberg}.} \bibinfo{year}{2024}\natexlab{}.
\newblock \showarticletitle{Same task, more tokens: the impact of input length on the reasoning performance of large language models}.
\newblock \bibinfo{journal}{\emph{arXiv preprint arXiv:2402.14848}} (\bibinfo{year}{2024}).
\newblock


\bibitem[Li et~al\mbox{.}(2024)]%
        {li2024translating}
\bibfield{author}{\bibinfo{person}{Ruishi Li}, \bibinfo{person}{Bo Wang}, \bibinfo{person}{Tianyu Li}, \bibinfo{person}{Prateek Saxena}, {and} \bibinfo{person}{Ashish Kundu}.} \bibinfo{year}{2024}\natexlab{}.
\newblock \showarticletitle{Translating c to rust: Lessons from a user study}.
\newblock \bibinfo{journal}{\emph{arXiv preprint arXiv:2411.14174}} (\bibinfo{year}{2024}).
\newblock


\bibitem[mark sweep(2025)]%
        {mark-sweep}
\bibfield{author}{\bibinfo{person}{mark sweep}.} \bibinfo{year}{2025}\natexlab{}.
\newblock \bibinfo{title}{mark-sweep}.
\newblock \bibinfo{howpublished}{\url{https://github.com/munificent/mark-sweep}}.
\newblock


\bibitem[mcrcon(2025)]%
        {mcrcon}
\bibfield{author}{\bibinfo{person}{mcrcon}.} \bibinfo{year}{2025}\natexlab{}.
\newblock \bibinfo{title}{mcrcon}.
\newblock \bibinfo{howpublished}{\url{https://github.com/Tiiffi/mcrcon}}.
\newblock


\bibitem[Nitin et~al\mbox{.}(2025a)]%
        {c2saferrust}
\bibfield{author}{\bibinfo{person}{Vikram Nitin}, \bibinfo{person}{Rahul Krishna}, \bibinfo{person}{Luiz~Lemos do Valle}, {and} \bibinfo{person}{Baishakhi Ray}.} \bibinfo{year}{2025}\natexlab{a}.
\newblock \bibinfo{title}{C2SaferRust: Transforming C Projects into Safer Rust with NeuroSymbolic Techniques}.
\newblock
\showeprint[arxiv]{2501.14257}~[cs.SE]
\urldef\tempurl%
\url{https://arxiv.org/abs/2501.14257}
\showURL{%
\tempurl}


\bibitem[Nitin et~al\mbox{.}(2025b)]%
        {nitin2025c2saferrust}
\bibfield{author}{\bibinfo{person}{Vikram Nitin}, \bibinfo{person}{Rahul Krishna}, \bibinfo{person}{Luiz Lemos~do Valle}, {and} \bibinfo{person}{Baishakhi Ray}.} \bibinfo{year}{2025}\natexlab{b}.
\newblock \showarticletitle{C2SaferRust: Transforming C Projects into Safer Rust with NeuroSymbolic Techniques}.
\newblock \bibinfo{journal}{\emph{arXiv preprint arXiv:2501.14257}} (\bibinfo{year}{2025}).
\newblock


\bibitem[Pan et~al\mbox{.}(2024)]%
        {pan2024lost}
\bibfield{author}{\bibinfo{person}{Rangeet Pan}, \bibinfo{person}{Ali~Reza Ibrahimzada}, \bibinfo{person}{Rahul Krishna}, \bibinfo{person}{Divya Sankar}, \bibinfo{person}{Lambert~Pouguem Wassi}, \bibinfo{person}{Michele Merler}, \bibinfo{person}{Boris Sobolev}, \bibinfo{person}{Raju Pavuluri}, \bibinfo{person}{Saurabh Sinha}, {and} \bibinfo{person}{Reyhaneh Jabbarvand}.} \bibinfo{year}{2024}\natexlab{}.
\newblock \showarticletitle{Lost in translation: A study of bugs introduced by large language models while translating code}. In \bibinfo{booktitle}{\emph{Proceedings of the IEEE/ACM 46th International Conference on Software Engineering}}. \bibinfo{pages}{1--13}.
\newblock


\bibitem[Pappas and Gazzillo(2024)]%
        {pappas2024semantic}
\bibfield{author}{\bibinfo{person}{Brent Pappas} {and} \bibinfo{person}{Paul Gazzillo}.} \bibinfo{year}{2024}\natexlab{}.
\newblock \showarticletitle{Semantic Analysis of Macro Usage for Portability}. In \bibinfo{booktitle}{\emph{Proceedings of the 46th IEEE/ACM International Conference on Software Engineering}}. \bibinfo{pages}{1--12}.
\newblock


\bibitem[Reference(2025)]%
        {cfg_attribute}
\bibfield{author}{\bibinfo{person}{The~Rust Reference}.} \bibinfo{year}{2025}\natexlab{}.
\newblock \bibinfo{title}{Conditional compilation}.
\newblock \bibinfo{howpublished}{\url{https://doc.rust-lang.org/reference/conditional-compilation.html }}.
\newblock


\bibitem[sds(2025)]%
        {sds}
\bibfield{author}{\bibinfo{person}{sds}.} \bibinfo{year}{2025}\natexlab{}.
\newblock \bibinfo{title}{sds}.
\newblock \bibinfo{howpublished}{\url{https://github.com/antirez/sds}}.
\newblock


\bibitem[Shetty et~al\mbox{.}(2024)]%
        {shetty2024syzygy}
\bibfield{author}{\bibinfo{person}{Manish Shetty}, \bibinfo{person}{Naman Jain}, \bibinfo{person}{Adwait Godbole}, \bibinfo{person}{Sanjit~A Seshia}, {and} \bibinfo{person}{Koushik Sen}.} \bibinfo{year}{2024}\natexlab{}.
\newblock \showarticletitle{Syzygy: Dual Code-Test C to (safe) Rust Translation using LLMs and Dynamic Analysis}.
\newblock \bibinfo{journal}{\emph{arXiv preprint arXiv:2412.14234}} (\bibinfo{year}{2024}).
\newblock


\bibitem[Silva et~al\mbox{.}(2023)]%
        {silva2023repairllama}
\bibfield{author}{\bibinfo{person}{Andr{\'e} Silva}, \bibinfo{person}{Sen Fang}, {and} \bibinfo{person}{Martin Monperrus}.} \bibinfo{year}{2023}\natexlab{}.
\newblock \showarticletitle{Repairllama: Efficient representations and fine-tuned adapters for program repair}.
\newblock \bibinfo{journal}{\emph{arXiv preprint arXiv:2312.15698}} (\bibinfo{year}{2023}).
\newblock


\bibitem[SipHash(2025)]%
        {SipHash}
\bibfield{author}{\bibinfo{person}{SipHash}.} \bibinfo{year}{2025}\natexlab{}.
\newblock \bibinfo{title}{SipHash}.
\newblock \bibinfo{howpublished}{\url{https://github.com/veorq/SipHash}}.
\newblock


\bibitem[su~exec(2025)]%
        {su-exec}
\bibfield{author}{\bibinfo{person}{su exec}.} \bibinfo{year}{2025}\natexlab{}.
\newblock \bibinfo{title}{su-exec}.
\newblock \bibinfo{howpublished}{\url{https://www.anthropic.com/pricing\#anthropic-api}}.
\newblock


\bibitem[System(2025a)]%
        {autoconf}
\bibfield{author}{\bibinfo{person}{GNU~Operating System}.} \bibinfo{year}{2025}\natexlab{a}.
\newblock \bibinfo{title}{Autoconf}.
\newblock \bibinfo{howpublished}{\url{https://www.gnu.org/software/autoconf/}}.
\newblock


\bibitem[System(2025b)]%
        {automake}
\bibfield{author}{\bibinfo{person}{GNU~Operating System}.} \bibinfo{year}{2025}\natexlab{b}.
\newblock \bibinfo{title}{Automake}.
\newblock \bibinfo{howpublished}{\url{https://www.gnu.org/software/automake/}}.
\newblock


\bibitem[tiny AES-c(2025)]%
        {tiny-AES-c}
\bibfield{author}{\bibinfo{person}{tiny AES-c}.} \bibinfo{year}{2025}\natexlab{}.
\newblock \bibinfo{title}{tiny-AES-c}.
\newblock \bibinfo{howpublished}{\url{https://github.com/kokke/tiny-AES-c}}.
\newblock


\bibitem[Tinyhttpd(2025)]%
        {Tinyhttpd}
\bibfield{author}{\bibinfo{person}{Tinyhttpd}.} \bibinfo{year}{2025}\natexlab{}.
\newblock \bibinfo{title}{Tinyhttpd}.
\newblock \bibinfo{howpublished}{\url{https://github.com/EZLippi/Tinyhttpd}}.
\newblock


\bibitem[Xue et~al\mbox{.}({[n.\,d.]})]%
        {xueinterpretable}
\bibfield{author}{\bibinfo{person}{Min Xue}, \bibinfo{person}{Artur Andrzejak}, {and} \bibinfo{person}{Marla Leuther}.} \bibinfo{year}{[n.\,d.]}\natexlab{}.
\newblock \showarticletitle{An interpretable error correction method for enhancing code-to-code translation}. In \bibinfo{booktitle}{\emph{The Twelfth International Conference on Learning Representations}}.
\newblock


\bibitem[Yang et~al\mbox{.}(2024b)]%
        {yang2024vert}
\bibfield{author}{\bibinfo{person}{Aidan~ZH Yang}, \bibinfo{person}{Yoshiki Takashima}, \bibinfo{person}{Brandon Paulsen}, \bibinfo{person}{Josiah Dodds}, {and} \bibinfo{person}{Daniel Kroening}.} \bibinfo{year}{2024}\natexlab{b}.
\newblock \showarticletitle{Vert: Verified equivalent rust transpilation with few-shot learning}.
\newblock \bibinfo{journal}{\emph{arXiv preprint arXiv:2404.18852}} (\bibinfo{year}{2024}).
\newblock


\bibitem[Yang et~al\mbox{.}(2024a)]%
        {yang2024exploring}
\bibfield{author}{\bibinfo{person}{Zhen Yang}, \bibinfo{person}{Fang Liu}, \bibinfo{person}{Zhongxing Yu}, \bibinfo{person}{Jacky~Wai Keung}, \bibinfo{person}{Jia Li}, \bibinfo{person}{Shuo Liu}, \bibinfo{person}{Yifan Hong}, \bibinfo{person}{Xiaoxue Ma}, \bibinfo{person}{Zhi Jin}, {and} \bibinfo{person}{Ge Li}.} \bibinfo{year}{2024}\natexlab{a}.
\newblock \showarticletitle{Exploring and unleashing the power of large language models in automated code translation}.
\newblock \bibinfo{journal}{\emph{Proceedings of the ACM on Software Engineering}} \bibinfo{volume}{1}, \bibinfo{number}{FSE} (\bibinfo{year}{2024}), \bibinfo{pages}{1585--1608}.
\newblock


\bibitem[yank(2025)]%
        {yank}
\bibfield{author}{\bibinfo{person}{yank}.} \bibinfo{year}{2025}\natexlab{}.
\newblock \bibinfo{title}{yank}.
\newblock \bibinfo{howpublished}{\url{https://github.com/mptre/yank}}.
\newblock


\bibitem[Yin et~al\mbox{.}(2024)]%
        {yin2024rectifier}
\bibfield{author}{\bibinfo{person}{Xin Yin}, \bibinfo{person}{Chao Ni}, \bibinfo{person}{Tien~N Nguyen}, \bibinfo{person}{Shaohua Wang}, {and} \bibinfo{person}{Xiaohu Yang}.} \bibinfo{year}{2024}\natexlab{}.
\newblock \showarticletitle{Rectifier: Code Translation with Corrector via LLMs}.
\newblock \bibinfo{journal}{\emph{arXiv preprint arXiv:2407.07472}} (\bibinfo{year}{2024}).
\newblock


\bibitem[Zhang et~al\mbox{.}(2025)]%
        {zhang2025scalable}
\bibfield{author}{\bibinfo{person}{Hanliang Zhang}, \bibinfo{person}{Cristina David}, \bibinfo{person}{Meng Wang}, \bibinfo{person}{Brandon Paulsen}, {and} \bibinfo{person}{Daniel Kroening}.} \bibinfo{year}{2025}\natexlab{}.
\newblock \showarticletitle{Scalable, validated code translation of entire projects using large language models}.
\newblock \bibinfo{journal}{\emph{Proceedings of the ACM on Programming Languages}} \bibinfo{volume}{9}, \bibinfo{number}{PLDI} (\bibinfo{year}{2025}), \bibinfo{pages}{1616--1641}.
\newblock


\bibitem[Zhang et~al\mbox{.}(2023)]%
        {zhang2023ownership}
\bibfield{author}{\bibinfo{person}{Hanliang Zhang}, \bibinfo{person}{Cristina David}, \bibinfo{person}{Yijun Yu}, {and} \bibinfo{person}{Meng Wang}.} \bibinfo{year}{2023}\natexlab{}.
\newblock \showarticletitle{Ownership guided C to Rust translation}. In \bibinfo{booktitle}{\emph{International Conference on Computer Aided Verification}}. Springer, \bibinfo{pages}{459--482}.
\newblock


\bibitem[Zhang et~al\mbox{.}(2024)]%
        {zhang2024systematic}
\bibfield{author}{\bibinfo{person}{Quanjun Zhang}, \bibinfo{person}{Chunrong Fang}, \bibinfo{person}{Yang Xie}, \bibinfo{person}{YuXiang Ma}, \bibinfo{person}{Weisong Sun}, {and} \bibinfo{person}{Yun Yang~Zhenyu Chen}.} \bibinfo{year}{2024}\natexlab{}.
\newblock \showarticletitle{A Systematic Literature Review on Large Language Models for Automated Program Repair}.
\newblock \bibinfo{journal}{\emph{arXiv preprint arXiv:2405.01466}} (\bibinfo{year}{2024}).
\newblock


\end{thebibliography}

% \appendix
% \input{11_memo}

% \newpage
% \todo{move some texts here, which will be removed later}
% \input{backup.tex}

% Yinzhi: Appendix is part of the 10 pages.  So we cannot include any. 

\end{document}